\documentclass{rstransa_jlt}

\usepackage[utf8]{inputenc}
\usepackage{hyperref}  
\usepackage{amsmath}
\usepackage{amsfonts}
\usepackage{amsthm}
\usepackage{amssymb}
\usepackage{graphicx}
\usepackage{color}
\usepackage{cite}
\usepackage{bm}

\theoremstyle{definition}

\theoremstyle{definition}

\newcommand{\nofrac}[2]{#1/#2}

\title{Open-flow mixing and transfer operators}
\author{Anna Kl\"unker$^{1}$, Kathrin Padberg-Gehle$^{1}$, Jean-Luc Thiffeault$^{2}$}
\address{%
  $^{1}$Leuphana Universit\"{a}t L\"{u}neburg,
  Institute of Mathematics and its Didactics, Applied Mathematics Group,
  L\"{u}neburg, Germany\\
  $^{2}$Department of Mathematics,
  University of Wisconsin -- Madison, Madison, WI, USA}

\corres{Jean-Luc Thiffeault\\
\email{jeanluc@math.wisc.edu}}

\subject{%
  76F25, 
  37N10, 
  37M25 
}


\begin{document}

\begin{abstract}
We study finite-time mixing in time-periodic open flow systems. We describe the transport of densities in terms of a transfer operator, which is represented by the transition matrix of a finite-state Markov chain. The transport processes in the open system are organized by the chaotic saddle and its stable and unstable manifolds.  We extract these structures directly from leading eigenvectors of the transition matrix. We use different measures to quantify the degree of mixing and show that they give consistent results in parameter studies of two model systems.
\end{abstract}

\begin{fmtext} 

\noindent
\textsl{Dedicated to the memory of Charlies R.\ Doering}

\section{Introduction}
\label{sec:intro}

In his pioneering work Aref \cite{Aref1984} introduced the concept of chaotic advection as the dynamical mechanism underlying stirring and mixing of scalars by fluid flows. For more than three decades the dynamical systems perspective on advective processes has inspired a multitude of scientific works (see the reviews \cite{Aref2017frontiers,DoeringNobili2017,thiffeault2012using}). In a closed chaotic flow, any blob of dye becomes increasingly filamentary under continuous stretching and folding. This advective process is accompanied by molecular diffusion, which by itself is a very slow process, but is enhanced by stirring to effectively mix the filaments homogeneously. Different mixing measures have been introduced \cite{danckwerts1952definition,mathew2005multiscale,kukukova2011measuring,lin2011,thiffeault2012using} and used as the basis for designing optimal stirring protocols \cite{mathew2007optimal,lin2011,thiffeault2012using,sinaober2015}; see also \cite{froyland2016optimal,froyland2017optimal,grover2018optimal,Marcotte2018b,Vermach2018} for recent relevant work on optimal mixing.

\end{fmtext} 

\maketitle

\begin{figure}
\begin{center}
\includegraphics[width=0.6\textwidth]{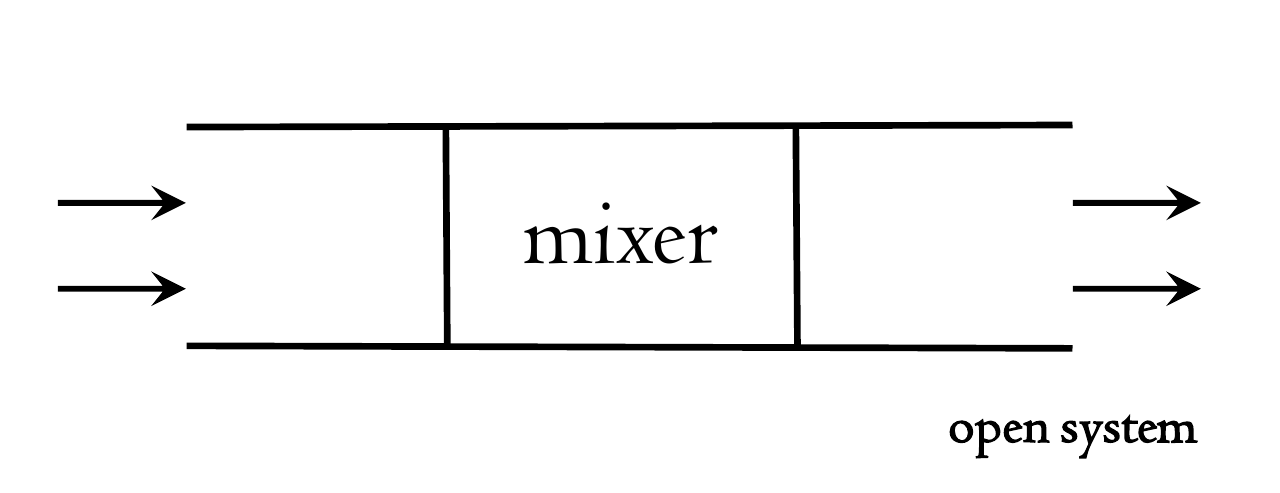}\\
\caption{Schematic representation of the model setup.\label{fig:scheme}}
\end{center}
\end{figure}

While much research has focused on the study of closed chaotic systems, in many application areas one has to deal with open flows, characterized by constant in- and outflows; such systems will be the focus of our work. Notable examples include the flow past an obstacle or the study of oil spills and algal blooms in the ocean. In such open systems, chaotic advection proceeds differently from the classical closed setting, as typical fluid particles leave any given region in finite time \cite{lai2011,moura2012}. Some rare fluid particles may stay forever and have trapped orbits confined to a fractal structure -- the chaotic saddle \cite{Jung1993,Neufeld1998,Neufeld2000b,Gouillart2009,Gouillart2011}. Fluid particles are asymptotically attracted to the chaotic saddle along its stable manifold, while attraction under time reversal is captured by the unstable manifold.
Although these structures have Lebesgue measure zero, they influence the mixing properties of the flow tremendously: fluid particles near the stable manifold are transported to a neighborhood of the chaotic saddle, where they experience repeated stretching and folding before they flow out along the unstable manifold. Unlike in closed flows, the resulting filamentary structures do not fill space. Thus the outgoing fluid is only partially mixed \cite{thiffeault2011}.

A suitable numerical framework for studying transport and mixing in dynamical systems is based on the spectral theory of transfer operators, as it captures the advective or advective-diffusive evolution of densities under the action of the underlying flow. Transfer operators are Markov operators and thus their spectrum lies in the unit circle. For steady closed flows the eigenfunction with unit eigenvalue corresponds to the stationary distribution. The rate of mixing is then quantified by the second largest eigenvalue, which controls the exponential rate at which initial distributions approach the stationary distribution. In the limit of small diffusivity these eigenfunctions are related to so-called strange eigenmodes \cite{Pierrehumbert1994,Rothstein1999,Voth2003,Liu2004,Thiffeault2004b,Haynes2005}, which have also been studied for open flows \cite{Gouillart2009, Gouillart2011}.

For steady closed flows a numerical approximation of the transfer operator is obtained via Ulam's method \cite{UlamS}. The resulting numerical operator can be interpreted as the transition matrix of a finite-state Markov chain. The evolution of densities under the underlying nonlinear flow can therefore be approximated by matrix and vector products. In particular, the stationary density is given as the eigenvector to the unit eigenvalue. For more details on transfer operator methods for transport and mixing, we refer to e.g.\ \cite{Dellnitz1999,Dellnitz2005,froyland2010,froyland2013,froyland2014almost}. In these works, as well as for the related mapping method \cite{Singh2009,Gorodetskyi2012}, the focus has been on transport processes involving mainly closed flows.

In the present paper, we propose a computational approach based on transfer operators that allows us to study and quantify mixing in specific models of open flows. We consider simple model settings, including the well-studied periodically-forced double gyre flow \cite{shadden2005definition} and the lid-driven cavity flow \cite{grover2012topological}, which are known to exhibit extended regions of stretching and folding. We combine these initially closed systems with a background flow that transports fluid particles through a compact phase space region where the periodic mixer acts (Fig.~\ref{fig:scheme}). Our specific setting, with a laminar flow in a channel being subjected to a periodic mixing protocol in a compact mixing region, is inspired by mixing devices that can be realized experimentally \cite{Gouillart2009,Gouillart2011,Ward_2015}. We describe the evolution of densities by means of a substochastic transition matrix, resulting from a finite-rank approximation of a Perron-Frobenius operator for the open system.
Additionally, the proposed framework allows us to visualize the crucial structures relevant for mixing, such as the chaotic saddle and its invariant manifolds, by means of leading eigenvectors of the transition matrix. This means that the dynamical organizers of mixing can be obtained from the numerical transfer operator with negligible additional computational effort.   In the same setup, we also study and quantify how two species of fluid -- with a given inflow distribution -- are mixed in such a system. In particular, the resulting outflow pattern corresponding to the partially-mixed fluid can be described as the stationary distribution of an affine transformation involving the numerical transfer operator as described by \cite{floriani2016flux}; see also \cite{berthiaux2004applications} for related Markov chain models of mixing processes. The partially-mixed fluid can then be analyzed by means of different established mixing measures; we study the system's mixing behavior for different parameters.

The paper is organized as follows:
In section~\ref{sec:setup_and_transport} we introduce the basic modeling setup and describe the transport of densities in closed and open flows within a transfer operator approach. We review the numerical approximation of the transfer operator, which results in a time-homogeneous finite-state Markov chain. The underlying dynamics and mixing processes can then be studied in terms of the flux through this Markov chain. In section~\ref{sec:quantification} we discuss several measures for the quantification of mixing, such as the variance or multiscale mix-norm. We also discuss the organizers of open flow mixing, such as the chaotic saddle, and their numerical approximation. In section~\ref{sec:examples} we study two sample systems, where a periodically forced closed system is combined with a constant background flow.
We conclude the paper with a discussion in section~\ref{sec:discussion}.

\section{Transport of densities}
\label{sec:setup_and_transport}
In this section, we address advective transport in open flows and introduce our specific mathematical model setup. For simplicity, we assume that the flow consists of ideal fluid particles that move passively following a given fluid velocity $\bm{u}(\bm{x},t)$ in a domain~$\bm{x} = (x,y) \in X \subset \mathbb{R}^2$. The particle trajectories obey the ODE
\begin{equation}
  \dot{\bm{x}}(t)=\bm{u}(\bm{x}(t),t),
  \qquad
  \bm{x}(0) = \bm{x}_0.
  \label{eq:ODEs}
\end{equation}
We restrict to incompressible ($\nabla\cdot\bm{u}=0$) time-periodic flows with period $\tau$ ($\bm{u}(\bm{x},t)= \bm{u}(\bm{x},t+\tau)$). We can then define an area-preserving stroboscopic flow map $T: X \to X$ with $T(\bm{x}_0)=\bm{x}(\tau)$, $T^2(\bm{x}_0)=\bm{x}(2\tau)$, etc.

\subsection{Transport and transfer operators in closed systems}

In studying transport and mixing we are concerned with the evolution of densities rather than single particles. Suppose that $T:M \to M$, where $M \subset X$ is compact. We can extend $M$ to a probability space $(M, \mathcal{B}(M), \mu)$, where $\mathcal{B}(M)$ is the Borel $\sigma$-algebra and $\mu$ is normalized Lebesgue measure, i.e. $\mu(B)=\nofrac{\ell(B)}{\ell(M)}$ for all $B\in \mathcal{B}(M)$, where $\ell$ denotes Lebesgue measure (area) on $\mathbb{R}^2$. As the flow map $T:M \to M$ is area-preserving, it holds that $\mu(B)=\mu(T^{-1}(B))$ for all $B \in \mathcal{B}(M)$, where $T^{-1}(\cdot)$ is the preimage under $T$ (i.e., the time-reversed flow map). The evolution of a probability density $f\in L^1(M,\mu)$ under the flow map $T$ can
be described by a linear operator, the Perron-Frobenius operator $\bar{\mathcal{P}}: L^1(M) \to L^1(M)$ \cite{Las}, defined by
\begin{equation}
  \int_{B} \bar{\mathcal{P}} f d \mu = \int_{T^{-1} B} f d \mu \quad  \text{for all } B\in \mathcal{B}(M).
\end{equation}
The Perron-Frobenius operator is a Markov operator. Here, $\bar{\mathcal{P}}\bm{1}_M=\bm{1}_M$ due to area-preservation of $T$ and thus $\|\bar{\mathcal{P}}\bm{1}_M\|=1$, where $\|\cdot \|$ denotes the normalized $L^1$ norm.

Using Ulam's method \cite{UlamS} we obtain a finite-rank approximation of the infinite-dimensional transfer operator $\bar{\mathcal{P}}$ as a matrix $\bar{P}$. To this end, the domain $M$ is partitioned into $n$ sets $B_i$, $i=1,\ldots, n$, where $B_i \cap B_j =\emptyset$ for all $i\neq j$. For implementation purposes both the domain $M$ and the partition elements $B_i$ are chosen to be rectangles, which we call boxes in the following.

The proportion of mass that is mapped from box $B_i$ to $B_j$ under $T$ is given by the matrix entry
\begin{equation}
  \bar{P}_{ij}=\frac{\mu(B_i \cap T^{-1}(B_j))}{\mu(B_i)}=\frac{\ell(B_i \cap T^{-1}(B_j))}{\ell(B_i)}.
\end{equation}
In practice, uniformly distributed test particles are initialized in box $B_i$. The entry $\bar{P}_{ij}$ of the matrix $\bar{P}$ is then estimated as the proportion of these particles that gets mapped to box $B_j$ under the action of $T$. We use the software package GAIO \cite{dellnitz2001algorithms} for the approximation of the transfer operator. The matrix $\bar{P}$ is row-stochastic and can be interpreted as the transition matrix of a time-homogeneous Markov chain on a finite state space, where the boxes are the states.

In the discrete context, densities are now represented by discrete probability measures $\nu$. The interpretation of the $\bar{P}$-induced dynamics is that if $\bm{v}_0\ge 0$ (component-wise) is a probability vector ($v_{0,i}=\nu(B_i)$, $\sum_i v_{0,i}=1$), then
$$\bm{v}_1=\bm{v}_0\bar{P}$$
is the push-forward of $\bm{v}_0$ under the discretized action of $T$. This extends to the evolution of any vector (e.g. signed densities) -- similar in spirit to the mapping method \cite{Singh2009,Gorodetskyi2012}.

\subsection{Transport in open flow systems}
\label{sec:setup_doublegyre}
\begin{figure}
\begin{center}
\includegraphics[width=0.99\textwidth]{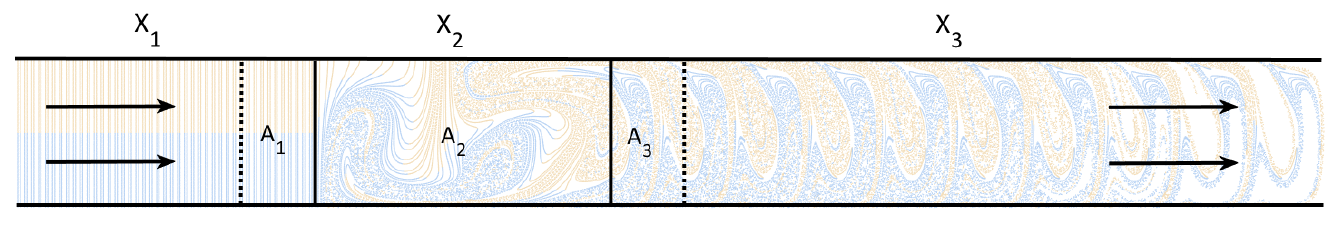} \\
\caption{Setup of the double gyre mixer.}
\label{fig:setup_doublegyre}
\end{center}
\end{figure}

Now, we formally consider the dynamics of a flow map $T: X \to X$ that is defined on an infinite strip $X \subset \mathbb{R}^2$ (Fig.~\ref{fig:setup_doublegyre}). We assume that $X$ can be divided into three regions: a background flow transports particles from an unbounded unmixed region $X_1$ into an unbounded mixed region $X_3$ after passing a bounded mixing region $X_2$.
When we send fluid with two different colors through the mixing region, in the mixed region a ``periodic" pattern is formed after some time and advected downstream by the background velocity field.

We want to describe the mass evolution on an open subsystem with domain $A$, where $A$ contains an inlet flow region $A_{1}\subset X_1$ and an outlet flow region $A_{3}\subset X_3$ as well as the mixing region $A_{2}=X_2$.
We define our compact, connected domain $A=\overline{A_{1} \cup A_{2} \cup A_{3}}$ in the following way.  Particles in the unmixed region $A_{1} \subset X_1$ enter the mixing region~$A_2$ in one time step $\tau$.  The set $A_{3} \subset X_3$ is then chosen to be of the same size as $A_{1}$ (Fig.~\ref{fig:setup_doublegyre}).

Let $T_A: A \to X$ be the restriction of $T$ on $A \subset X$ and let $\mu_A$ be normalized Lebesgue measure on $A$, i.e. $\mu_A(B)=\nofrac{\ell(B)}{\ell(A)}$ for all $B \in \mathcal{B}(A)$.
We can formally write $T_A:(A, \mathcal{B}(A), \mu_A) \to (X,  \mathcal{B}(X), \mu_A)$. By construction $A$ is not invariant under the action of the flow map, i.e. $T_A(A) \not\subset  A$ and $T_A^{-1}(A)  \not\subset A$, as some mass leaves $A$ under iteration of $T_A$ and~$T_A^{-1}$.
Analogously to the closed case, we can define a conditional transfer operator
\begin{equation}
  \mathcal{P}: L^1(A) \to L^1(A),
\end{equation}
via
 \begin{equation}
  \int_{B} \mathcal{P} f d \mu_A = \int_{T^{-1} B} f d \mu_A \quad  \text{for all } B\in \mathcal{B}(A).
\end{equation}
Due to the leakage out of~$A$, we have $\|\mathcal{P} \bm{1}_A\|<1$.

The numerical approximation of $\mathcal{P}$ in terms of a transition matrix $P$ can be derived in a similar manner as in the closed case. Let now the set $A$ be partitioned into $n$ sets $B_i$, $i=1,\ldots, n$, where $B_i \cap B_j =\emptyset$ for all $i\neq j$ and form the matrix
\begin{align}
  P_{ij}
  = \frac{\mu_A(B_i \cap T_A^{-1}(B_j))}{\mu_A(B_i)}
  = \frac{\mu_A(B_i \cap T^{-1}(B_j))}{\mu_A(B_i)}
  &= \frac{\ell(B_i \cap T^{-1}(B_j))}{\ell(B_i)}.
\end{align}
 The evolution of a (signed) density can be again represented by the discrete push-forward of a vector $\bm{v}_0 \in \mathbb{R}^n$ via  $\bm{v}_1=\bm{v}_0P$.
 Unlike $\bar{P}$ in the closed case, $P$ is no longer row-stochastic but row-substochastic as there is constant outflow of mass. Under the assumption that all particles can (eventually) leave $A$, the matrix $P$ can be considered as the transition matrix of an absorbing Markov chain restricted to finite transient states.

\subsection{Flux through a Markov chain}
\label{sec:markov_flux}

In the following, we consider the Markov chain obtained from a finite-rank approximation of the transfer operator as described above. In particular, we study the evolution of a mass distribution in an open system with constant inflow and outflow of mass in terms of a Markov chain \cite{floriani2016flux}.

To this end, we first look at a closed system in terms of a time-homogeneous absorbing Markov chain on the state space $B=\{B_i,\, i=1,\ldots, N\}$. Let the $N \times N$ transition matrix $\bar{P}$ contain the transition probabilities between the states for a discrete time step of length $\tau$.
A Markov chain is \textit{time-homogeneous} if the transition matrix $\bar{P}$ is the same for each time step. A time-homogeneous transition matrix is obtained when approximating the transfer operator for a steady flow or, as in our setting, the stroboscopic map of a time periodic flow.
A state $B_i$ is said to be \textit{transient} if, given that we start in state $B_i$, there is a non-zero probability that we never return to $B_i$. A state $B_i$ is called an \textit{absorbing state} or \textit{sink} if $\bar{P}_{ij}=0$ for all $j\neq i$, so in this state mass is accumulated. A Markov chain is \textit{absorbing} if every state eventually reaches an absorbing state (mass flows from the nonabsorbing states to the sinks). All states that are nonabsorbing are then transient. In our set-up (Fig.~\ref{fig:setup_doublegyre}), we model states that are not on the domain $A$ of the open system as finite absorbing states, since we are not interested in them. Choosing the length of the time steps as the time period of the mixer results in a time-homogeneous Markov chain.

Let now $\mathcal{B}_{\mathrm{t}}=\{B_i, i=1,\ldots, n\}$ be the set of transient states and let $\mathcal{B}_{\mathrm{a}}=\{B_i, i=n+1,\ldots, N\}$ be the set of absorbing states. Let $P$ be the $n\times n$ submatrix of the transition matrix $\bar{P}$ obtained by removing the rows and columns of the $N-n$ absorbing states. Then ${P}$ is substochastic and describes the mass evolution on a system with outflow, just like the transition matrix of the open system $T_A: A \to X$ introduced in the previous section.

Since $A$ is a subset of an infinite strip $X$, we also want to take into account new mass that comes into our domain $A$. The constant source is modeled via the mass distribution ${\bm\sigma}$ over $\mathcal{B}_{\mathrm{t}}$, that is injected in the system at every time step $\tau$. A state $B_i$ is called called \textit{source state} if ${\sigma}_i \neq 0$. In our set-up, we model boxes that are on $A_{1}$ as source states.

Now let $\bm{v}_k$ denote the mass distribution restricted to the nonabsorbing states after $k\in \mathbb{Z}$ time steps. The mass distribution $\bm{v}_k$ is given by
\begin{equation}
  \bm{v}_k= \bm{v}_{k-1} {P}+ \bm{\sigma}
\end{equation}
or -- starting with the initial distribution $\bm{v}_0$ -- by
\begin{equation}
  \bm{v}_k= \bm{v}_0 {P}^k + \bm{\sigma} ({P}^{k-1}+{P}^{k-2}+ \ldots + I).
\end{equation}
The term $\bm{v}_0 {P}^k$ tends to 0 and the geometric series converges. Hence the mass distribution converges to
\begin{equation}\label{eq:vinv}
  \bm{v}_{\text{inv}}:= \lim_{k\to\infty} \bm{v}_k=\bm{\sigma} (I-{P})^{-1}.
\end{equation}
The matrix $(I-{P})^{-1}$ is called the \textit{fundamental matrix}. The $(i,j)$th entry of the fundamental matrix can be interpreted as the expected number of visits of a particle in state $B_j$ when starting in $B_i$ (before leaving $A$).
Note that the invariant mass distribution  $\bm{v}_{\text{inv}}$ in \eqref{eq:vinv} is independent of the initial mass distribution $\bm{v}_0$, but  depends on $\bm{\sigma}$, which is injected into the system after each time step.

\section{Mixing in open flows}
\label{sec:quantification}

Mixing refers to ``the intermingling of distinct materials or fluid properties that were originally separated in space'' \cite{DoeringNobili2017}. In realistic flows, molecular diffusion homogenizes an initially inhomogeneous spatial distribution of a scalar tracer, but this is a very slow process. When attempting to accelerate homogenization, i.e. maximizing the rate of mixing, advection is required in order to create small-scale structures in the spatial distribution of the advected fields, which are then quickly smoothed by diffusion.  While turbulence is the classical mechanism, we consider chaotic advection, where the corresponding small-scale structures are created by means of the stretching and folding property of chaotic flows~\cite{Aref1984}. By this mechanism the length scales of the resulting structure in the contracting direction decrease exponentially fast, and thus will be smoothed out by diffusion. This is a purely kinematic effect. Mixing by chaotic advection does not require a high Reynolds number and thus can be set up also in microfluidics \cite{Aref2017frontiers}.

Mixing by stretching and folding properties of the underlying flow relates to the classical ergodic theoretical concept of mixing~\cite{Lasota}. To see this, let $(X,\mathcal{B}(X), \mu)$ be a probability space and $T: X \to X$ a measure-preserving transformation (e.g.\ the flow map of the ODEs~\eqref{eq:ODEs} with $\mu$ being normalized Lebesgue measure). $T$ is called mixing if
\begin{equation}
  \lim_{k\to \infty} \mu(M_1\cap T^{-k}(M_2)) = \mu(M_1)\,\mu(M_2)  \quad  \text{for all } M_1,M_2 \in \mathcal{B}(X).
\end{equation}
For any two measurable subsets $M_1$ and $M_2$ the proportion of particles starting in $M_1$ and ending in $M_2$ after a long time is given asymptotically as the product of $\mu(M_1)$ and $\mu(M_2)$.  However, while this is an asymptotic concept, mixing in most technically-relevant systems can only be studied on a finite time span. So we are interested rather in the degree of mixing, which can be quantified by different measures of mixing as we outline below.

\subsection{Quantifying mixing}
In the following we review three different measures that can be used to quantify mixing in a setting with two types of fluid material.  These types have the same properties and are advected by the same velocity field, and differ only in ``color". This information is encoded in a scalar field $c(\cdot)$ on an outlet region $A_{3}$, which is preserved by the flow. $c(\cdot)$ has zero mean and its sign distinguishes the two differently-colored fluid materials. We want to measure the degree of mixing of this scalar field $c(\cdot)$.

For the domain discretized into $n$ boxes, the scalar field is given as a discrete $n$-vector $\bm{v}$, where an entry ${v}_i$ is the mass in box $B_i$. In our discrete model, we would consider a system perfectly mixed if in each box in the outlet the proportion of the two types is equal to the proportion of the two types in the source. This is in agreement with the ergodic theoretical definition of mixing.

Danckwerts \cite{danckwerts1952definition} considered the intensity and the scale of segregation to quantify mixing and we briefly discuss these measures in the following.

\paragraph{Sample variance.}
A classical measure of the intensity is the variance of the scalar field (or $L^2$-norm for a zero-mean field). We calculate the sample variance by
\begin{equation}
  s^2=\frac{1}{n-1}\sum_{i=1}^n ( {v}_i- \bar{{v}})^2,
\end{equation}
where $\bar{{v}}$ is the sample mean.
Note that in a purely advective setting, that is without diffusion, the variance of the scalar field is unaffected by the dynamics.
However, the transfer operator approach exhibits numerical diffusion such that the sample variance is a meaningful measure of advective mixing in our context.

\paragraph{Mean length scale.}
The semivariogram $\gamma(h)$ is a measure of spatial autocorrelation and provides information on the scales \cite{kukukova2011measuring}. Let $z_i$ be the indicator value in box $B_i$, $i=1, \ldots, n$, that is 1 if $v_i>0$ and 0 otherwise. We calculate the empirical indicator semivariogram by the estimator
\begin{equation}
  \hat{\gamma}(h)= \frac{1}{2 |N(h)|} \sum_{\substack{(B_i,B_j) \in N(h) }} (z_i -z_j)^2,
\end{equation}
where $(B_i,B_j) \in N(h)$, if the distance of the centers of $B_i$ and $B_j$ is $h\pm \delta$ for a bin width $2 \delta$. The inverse of the initial slope of the variogram gives the mean length scale \cite{carle1996transition}
\begin{equation}
  L=\frac{1}{2} \left[\frac{\partial\gamma}{\partial h}\right]_{h\to 0}^{-1}.
\end{equation}
In our parameter studies we estimate the mean length scale using the slope of the empirical indicator semivariogram  $\hat{\gamma}$ and take into account spatial periodicity.

\paragraph{Mix-norm.} A multiscale measure of mixing is the mix-norm
(equivalent to a negative Sobolev norm) introduced in
\cite{mathew2005multiscale}. In the sense of the mix-norm a scalar field is considered to be well-mixed if its averages over arbitrary open sets are uniform. The variance by itself does not capture this property. In contrast to the $L^2$-norm of a function, which is obtained by integrating the square of the function over the whole space, the mix-norm is obtained by integrating the square of average values of the function over a dense set of subsets contained in the whole space.

The mix-norm $\phi(c)$ of a scalar field $c(\cdot)$ on a 2-dimensional torus $T^2$ is defined as
\begin{equation}
  \phi^2(c)=\int_0^1 \int_{T^2} \left(\frac{\int_{\bm{x}\in B(\bm{p},s)} c(\bm{x}) \mu(d\bm{x})}{\mu (B(\bm{p},s))}\right)^2 \mu(d\bm{p}) \mu(ds),
\end{equation}
where $B(\bm{p},s)$ is a $2$-dimensional ball with center $\bm{p}$ and radius $s/2$.
In particular, $\phi^2(c)$ is the mix-variance for a zero-mean field $c$.
For a scalar field $c(\cdot)$ with Fourier expansion $c(\bm{x}) = \sum_{\bm{k}} c_{\bm{k}} e^{i2\pi(\bm{k}\cdot \bm{x})}$, the mix-norm $\phi(c)$ is given by
\begin{equation}
  \phi^2(c) = \sum_{\bm{k}} \frac{1}{(1+(2\pi \| \bm{k} \|)^2)^\frac{1}{2}}|c_{\bm{k}}|^2.
\end{equation}

For ease of use we calculate the mix-norm in our examples using the Fourier transformation. Although our domain is not a torus, it is closed in one coordinate direction and periodic in the other, and thus could be extended to a doubly-periodic domain.

\subsection{Organizers of mixing}

\paragraph{Chaotic saddle.}  A \textit{chaotic saddle} (or \textit{nonattracting chaotic invariant set}) $C \subset A$  is an invariant set of the open system $T_A : A\to X$ (i.e. $T_A(C)=C$) that contains a dense chaotic orbit. Trajectories in $C$ thus never escape the mixing region. The \textit{stable manifold} $S_C$ contains trapped trajectories that converge towards the chaotic saddle in forward time. The \textit{unstable manifold} $U_C$ contains trajectories that converge to the chaotic saddle in backward time. It follows that $S_C$ is forward invariant ($T_A(S_C) \subset S_C$) and $U_C$ is a backward invariant set ($T_A^{-1}(U_C) \subset U_C$).

While the chaotic saddle and its manifolds have Lebesgue measure zero, they still have a profound impact on the finite-time mixing properties of the flow. In particular, particles near the stable manifold of the chaotic saddle stay longer in the system and follow the unstable manifold on their way out. Therefore the geometry and fractal dimension of these structures has been studied and related to the mixing properties of the flow \cite{lai2011, Aref2017frontiers, Jung1993, ott2002chaos}. Open flow systems can also contain islands of positive Lebesgue measure (invariant sets that contain stable orbits) in the mixing region, where material is trapped forever and no material from outside of these islands gets inside.  We refer to the chaotic saddle, its manifolds, and to islands collectively as (invariant) phase-space structures.

\paragraph{Accessible states and classes.}  The invariant phase-space structures that organize mixing will be extracted directly from the $n \times n$ transition matrix  $P$, defined in section \ref{sec:setup_and_transport}. Recall that the region of interest $A\subset X$ is discretized by boxes $\{B_1, \ldots, B_n\}$, and $P_{ij}$ gives the conditional probability that a randomly selected point in $B_i$ has its image in $B_j$.

A state $B_j$ is \textit{accessible} or \textit{reachable} from state $B_i$ if there is a nonzero probability to reach $B_j$ after a finite number of steps when starting from $B_i$, i.e. $P^k_{ij}>0$ for some $k\in \mathbb{N}_0$. A state $B_i$ is said to \textit{communicate} with state $B_j$ if $B_j$ is accessible from $B_i$ and $B_i$ is accessible from $B_j$. A \textit{communicating class} is a maximal set of states $S$ such that every pair of states in $S$ communicates with each other. A communicating class $S_j$ is accessible from another communicating class $S_i$ if there is a state in  $S_j$ that is accessible from a state in $S_i$.

In the general case where there is more than one communicating class ($P$ is then reducible), after sorting the states appropriately by communicating class and reachability between the classes, the transition matrix takes the form of a block-lower-triangular matrix. In the sorted matrix the blocks $P_1, \ldots, P_r$ describe the transition probabilities restricted to single communicating classes $S_1, \ldots, S_r$ of the Markov chain and, consequently, the eigenvalues of the transition matrix $P$ are the eigenvalues of $P_1, \ldots, P_r$. Classes that are not self-communicating consist only of a single state with eigenvalue zero; all other classes are self-communicating. The non-zero part of the spectrum of $P$ is therefore determined by the spectrum of $P$ restricted to self-communicating classes \cite{van2009quasi}.

\paragraph{Reconstructing phase-space structures.}  We now relate the phase-space structures of interest, which are approximately represented by unions of boxes, to states and classes in the Markov chain defined in terms of the (substochastic) transition matrix $P$. It is clear that those sets of boxes in the discretization of the domain $A$ that correspond to self-communicating classes can trap particles (by definition there is a non-zero probability to stay in the respective set of boxes) and thus play a crucial role in mixing. In particular, they can either reveal a chaotic saddle or an invariant (or almost invariant) set with positive Lebesgue measure.
In practice, the self-communicating classes of $P$ can be found by graph-based approaches such as Tarjan's algorithm~\cite{Tarjan1972}.

In the following, we describe an alternative approach based on the spectrum of $P$. The set of particles that stay in $A$ for all times can be computed as $A_{\infty}=\bigcap_{k\in \mathbb{N}} T_A^k(A)$, which is a backward-invariant set \cite{DellnitzHohmann1997}. In the open flow literature \cite{lai2011, Aref2017frontiers, Jung1993, ott2002chaos}, $A_{\infty}$ corresponds to the unstable manifold when $A$ contains a chaotic saddle. Now let $\bm{w}_0=\frac{1}{n}\bm{1}$ and consider $\bm{w}_{k+1}=\bm{w}_{k}P$, where $\bm{w}_{k}$ is normalized at each step. Then it is intuitively clear that the limiting vector (i.e.\  $\bm{w}_{\infty}$, which could also be obtained in terms of an eigenvector of $P$) is supported on $A_{\infty}$. Similarly, when $P^T$ is considered, the limiting vector is supported on the stable manifold.

We now make this intuitive notion more precise using results from \cite{van2009quasi}.
With~$P$ sorted into blocks~$P_1,\ldots,P_r$ as above, let $\lambda_i$ denote the eigenvalue of $P_i$ with largest real part -- the \textit{Frobenius-Perron eigenvalue}.  The eigenvalue $\lambda_i$ is real and nonnegative. A class is called a \textit{maximal class} if $\lambda_j < \lambda_i$ for all $j\neq i$ such that $S_j$ is accessible from $S_i$.
There exists a nonnegative real left $\lambda_i$-eigenvector $\bm{w}$ of the substochastic transition matrix $P$ if and only if $S_i$ is a maximal class, where the states in the support of $\bm{w}$ are accessible from the respective class. A nonnegative left $\lambda_i$-eigenvector $\bm{w}$ is a linear combination of eigenvectors with this property corresponding to maximal classes with the same eigenvalue. The nonnegative left $\lambda_i$-eigenvectors $\bm{w}$ correspond to quasi-stationary distributions, i.e. $\bm{w}$ is invariant under time evolution conditioned on absorption not yet having taken place.

An analogous result holds for nonnegative right $\lambda_i$-eigenvectors $\bm{\hat{w}}$ of $P$. Here the respective classes are maximal classes of $P^T$ and are reachable by the states in the support of the right eigenvector.
In general, not all self-communicating classes are maximal, and one obtains signed leading eigenvectors for the respective non-maximal classes. However, a single self-communicating class, or several self-communicating classes that are not reachable from each other, correspond to maximal classes of $P$ and $P^T$.

When a maximal class $S_i$ with simple Frobenius-Perron eigenvalue $\lambda_i$ includes a chaotic saddle, states in the support of its left nonnegative eigenvector can be reached by $S_i$, and states in the support of the its right nonnegative eigenvector can reach $S_i$. In section~\ref{sec:examples} we therefore use the states in the support of the left (right) eigenvector to approximate the unstable (stable) manifold of the chaotic saddle. The intersection of the two invariant manifolds corresponds to the chaotic saddle.

Alternatively, the location of a stable manifold can be detected by high expected residence times in the mixing region. These can be calculated by means of the fundamental matrix~$(I-{P})^{-1}$. The expected number of time steps a particle is in a set of states $A$, when starting in a given state $B_i$, is
\begin{equation}
  \sum_{B_j\in A} (I-{P})^{-1}_{ij}.
\end{equation}

For the sake of completeness, we note that almost invariant sets, which are characterized as phase-space regions of positive Lebesgue measure that minimally mix with their surroundings, can also be considered as organizers of mixing. In closed systems, these can be identified by means of eigenvectors of $\bar{P}$ with real eigenvalue close to one \cite{Dellnitz1999,froyland2014almost}. The treatment of open systems, in the sense of having a small hole through which mass can escape, is considered in \cite{froyland2014closing}. In our context, these sets will be identified as self-communicating classes just like a chaotic saddle.

\section{Examples}
\label{sec:examples}

We consider now two sample systems on a domain $X \subset \mathbb{R}^2$ (Fig.~\ref{fig:setup_doublegyre}) as explained in section \ref{sec:setup_and_transport}\ref{sec:setup_doublegyre}. As the velocity field on $X$ is assumed to be divergence-free, it can be derived from a stream function $\Psi: X \times \mathbb{R} \to \mathbb{R}$, with
\begin{equation}
  \Psi(x,y,t)=\Psi_{\mathrm{b}}(x,y)+\Psi_{\mathrm{m}}(x,y,t).
\end{equation}
Here $\Psi_{\mathrm{b}}$ is the stream function of the stationary background flow, which shifts fluid through the mixing region and acts on all of~$X$, and $\Psi_{\mathrm{m}}$ is the stream function of the mixer, which acts only on $X_2$. The mixer is time-periodic: $\Psi_{\mathrm{m}}(x,y,t)=\Psi_{\mathrm{m}}(x,y,t+\tau)$, where $\tau>0$ is the the period. For a smooth superposition of the two stream functions $\Psi_{\mathrm{b}}$ and $\Psi_{\mathrm{m}}$, a standard bump function
supported on $X_2$ can be used (see Appendix~\ref{sec:bump}). The velocity field $\bm{u}=(u_1,u_2)$ on $X$ is then obtained from
\begin{equation}
u_1  =  \frac{\partial \Psi}{\partial y}, \qquad
u_2  =  -\frac{\partial \Psi}{\partial x},
\label{eq:ODEs2}
\end{equation}
%
%
and has the form
\begin{equation}
  \bm{u}(x,y,t)=\begin{cases} \bm{u}_{\mathrm{b}}(x,y),  & \text{for } (x,y) \in X_1 \cup X_3\\  \bm{u}_{\mathrm{b}}(x,y) + \bm{u}_{\mathrm{m}}(x,y,t), &\text{for } (x,y)\in X_2\end{cases}
\end{equation}
where $\bm{u}_{\mathrm{b}}$ is the stationary background velocity field and $\bm{u}_{\mathrm{m}}$ is the velocity field of the time-periodic mixer. We model the background flow as a constant homogeneous flow (translational flow), but other flow profiles such as parabolic flows are possible as well.

\subsection{The double gyre mixer}
As our first example we use the well-known periodically perturbed double gyre flow \cite{shadden2005definition} as the mixer. It has the stream function
\begin{equation}
  \Psi_{\mathrm{m}}= -\alpha \sin(f(x,t)\pi)\sin(\pi y),
\end{equation}
where $f(x,t)=\epsilon \sin(\omega t) x^2 +(1-2\epsilon \sin(\omega t)) x$ models the periodic perturbation with amplitude $\epsilon \geq 0$ and frequency $\omega$, and $\alpha>0$ controls the amplitude of the rotation speed of the gyres. We set $\omega=2 \pi$, so that the period is $\tau=1$.  We consider the domain $X=(-\infty,\infty) \times [0,1]$, with the unmixed region $X_1=(-\infty,0)\times[0,1]$, mixed region $X_3=(2,\infty)\times[0,1]$, and the mixing region $X_2=A_2=[0,2]\times[0,1]$ as discussed above.

\begin{figure}[!htb]
\begin{center}
\includegraphics[width=0.4\textwidth]{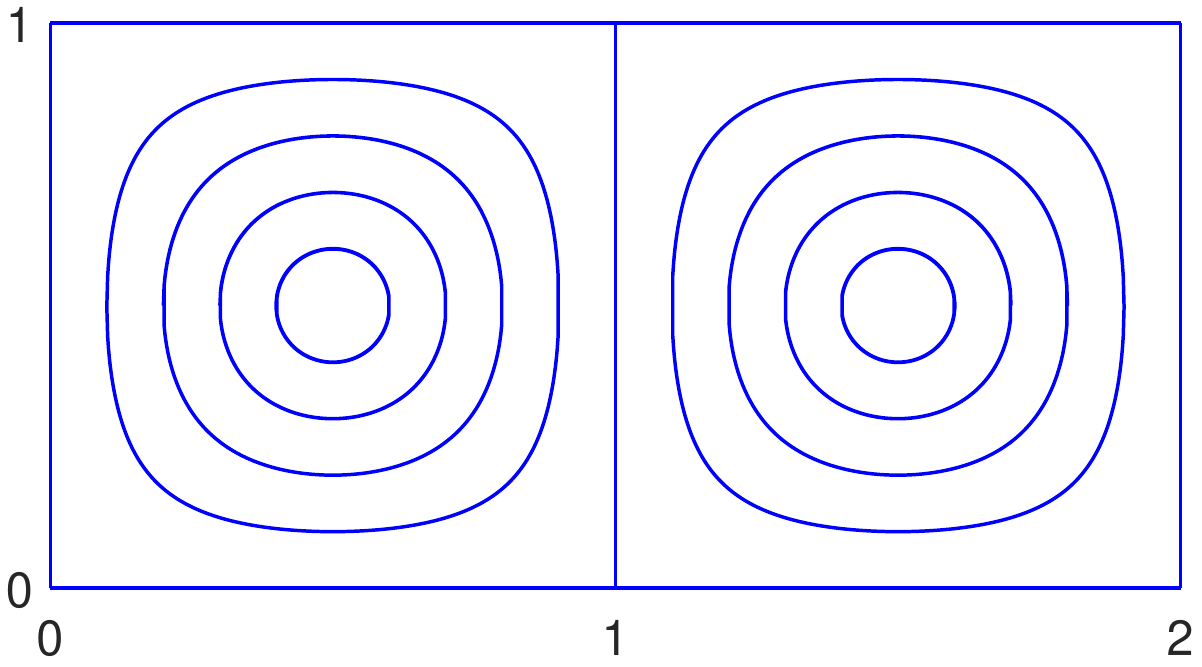}\hspace{2em}
\includegraphics[width=0.4\textwidth]{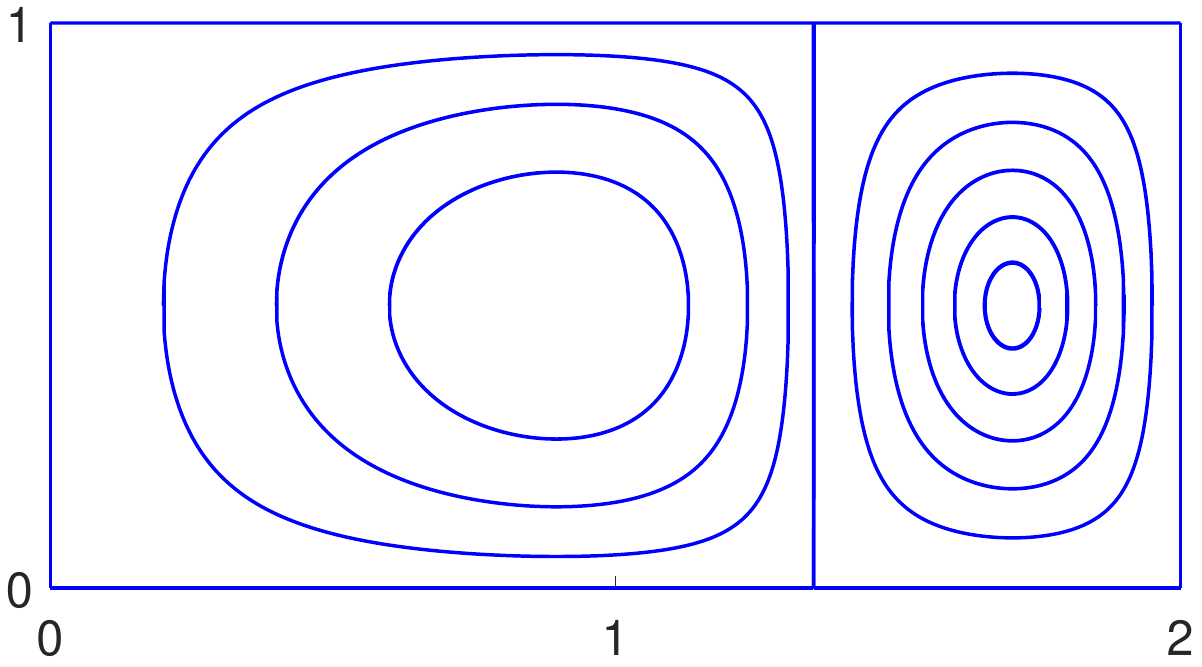}\\
\caption{Streamlines of the double gyre flow with $\alpha=0.5$ and  $\epsilon = 0.4$ at $t=0$ (left)  and $t=0.25$ (right).}\label{fig:doublegyre}
\end{center}
\end{figure}

For $\epsilon=0$, the flow is time-independent and the phase portrait displays two counter-rotating gyres separated by a heteroclinic orbit connecting the hyperbolic fixed points at $(1,0)$ and $(1,1)$. For $\epsilon>0$ the separatrix moves periodically. For~$\epsilon\ll 1$, $\epsilon$ is approximately the maximum displacement of the separation line from the middle to the left or right, reached at times 0.25 and 0.75, respectively. For large $\epsilon$ more than two gyres are formed in $X_2$ at times 0.25 and 0.75.  The streamlines of the double gyre flow on $X_2$ for two different time instances $t=0$ and $t=0.25$ are shown in Fig.~\ref{fig:doublegyre}, with parameters $\alpha=0.5$ and $\epsilon = 0.4$.
For the background flow, we choose the constant velocity field
\begin{equation}
  \Psi_{\mathrm{b}}= \beta y, \qquad
  \bm{u}_{\mathrm{b}}= \begin{pmatrix} \beta \\0\end{pmatrix}, \qquad
  \mbox{ with } \beta >0.
\end{equation}
The velocity field on $X$ has the form
\begin{equation}
  \bm{u}(x,y,t)=\bm{u}_{\mathrm{b}}(x,y)+\bm{u}_{\mathrm{m}}(x,y,t) \bm{1}_{[0,2]}(x),
\end{equation}
where $\bm{u}_{\mathrm{m}}(x,y,t)$ is derived from the stream function $\Psi_{\mathrm{m}}$.

We choose the inlet region $A_{1}=[-\beta,0) \times [0,1]$, which ensures that all mass is carried into the mixing region $A_{2}$ within a unit time step, the period of the mixer.  For the outlet region, we choose $A_{3}=(2,2+\beta] \times [0,1]$.
For setting up the finite state Markov chain, we divide the domain $A=[-\beta,2+\beta]\times [0, 1]$ in $49,152$ square boxes of side length $0.0078$ and initialize 100 test particles on a regular grid in each box. To form the transition matrix restricted to non-absorbing states, we integrate the test particles with the classical Runge-Kutta method with step size $h=0.01$ from $t=0$ to $t=1$, which corresponds to one period of the periodic mixer dynamics.
The boxes in $A_{1}$ correspond to the source states. As a constant source we consider a signed mass distribution describing the two different types (or colors) of fluid. Boxes with particles of the first type get the value~$1$ and boxes with the second type~$-1$.

\begin{figure}[!htb]
\begin{center}
\begin{tabular}{cc}
  \includegraphics[width=0.45\textwidth]{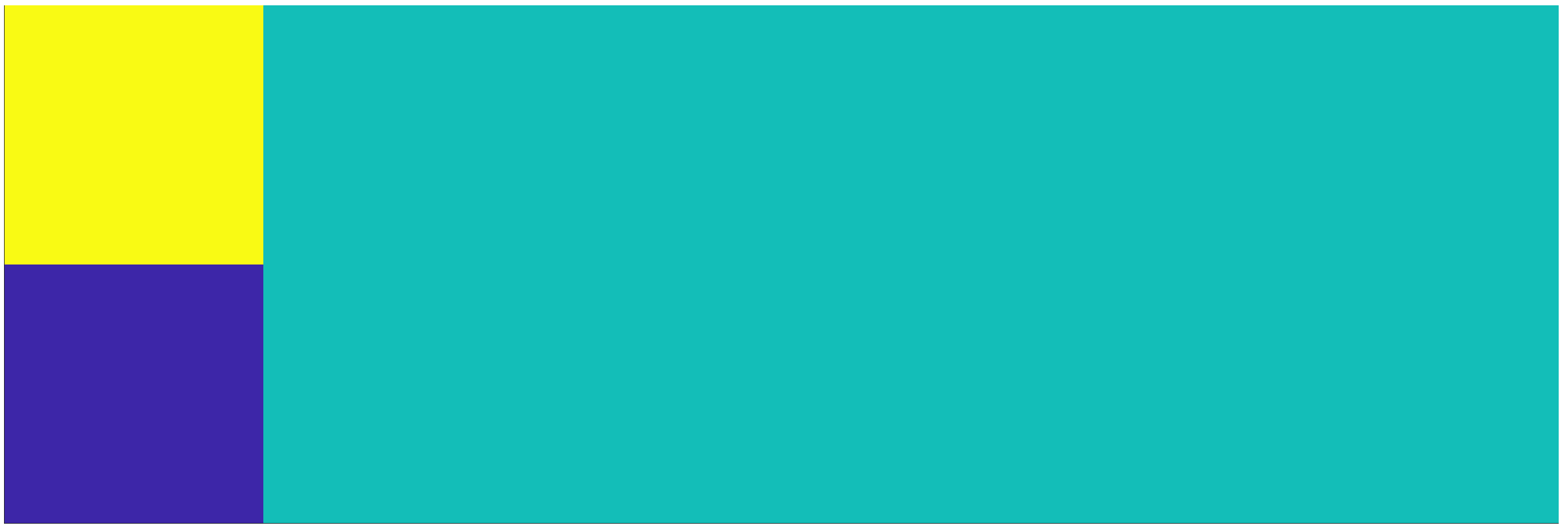} &
  \includegraphics[width=0.45\textwidth]{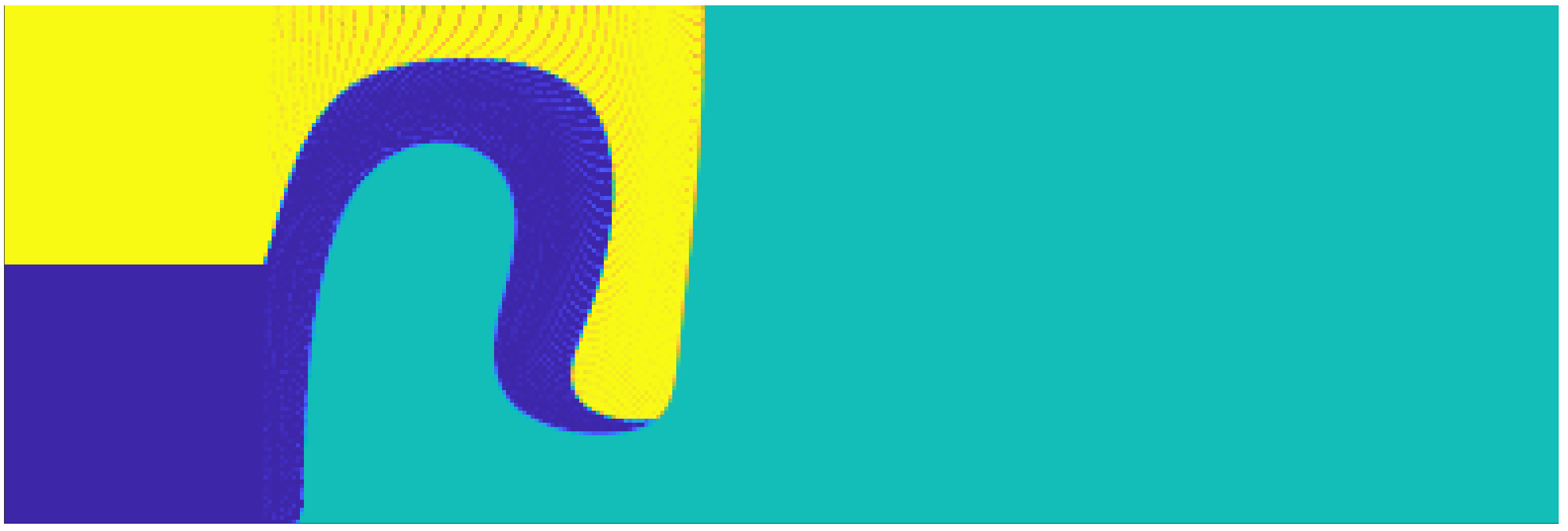} \\
  {\scriptsize $\bm{v}_1$} & {\scriptsize $\bm{v}_2$} \\[2mm]
  \includegraphics[width=0.45\textwidth]{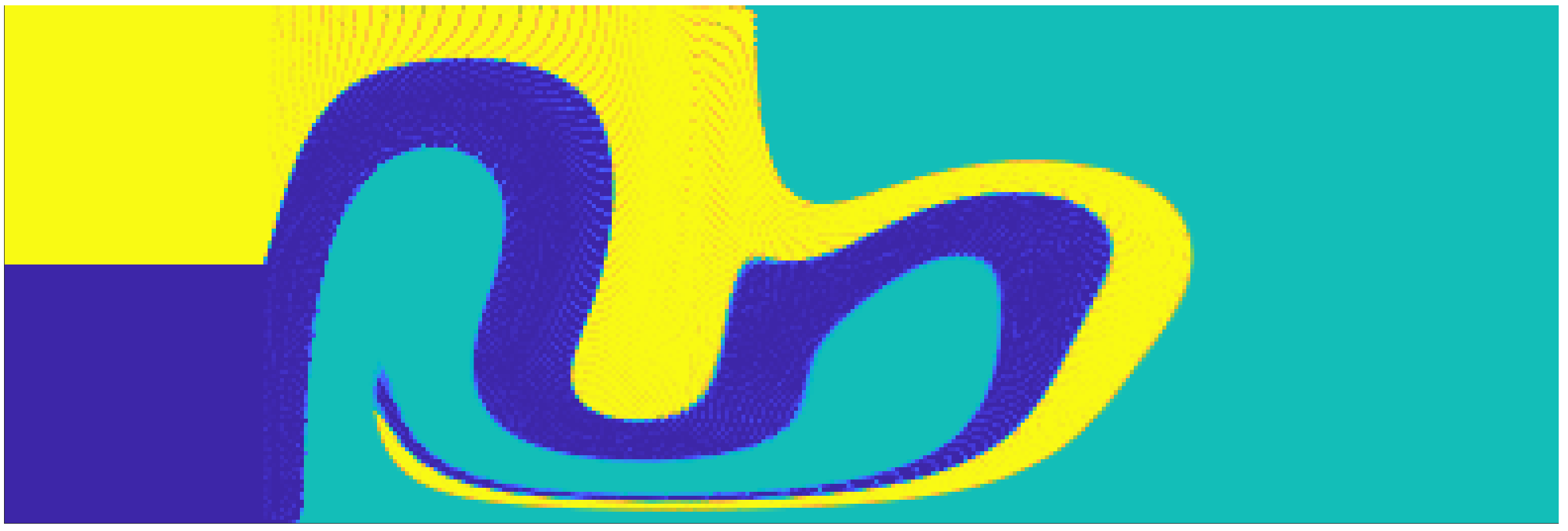} &
  \includegraphics[width=0.45\textwidth]{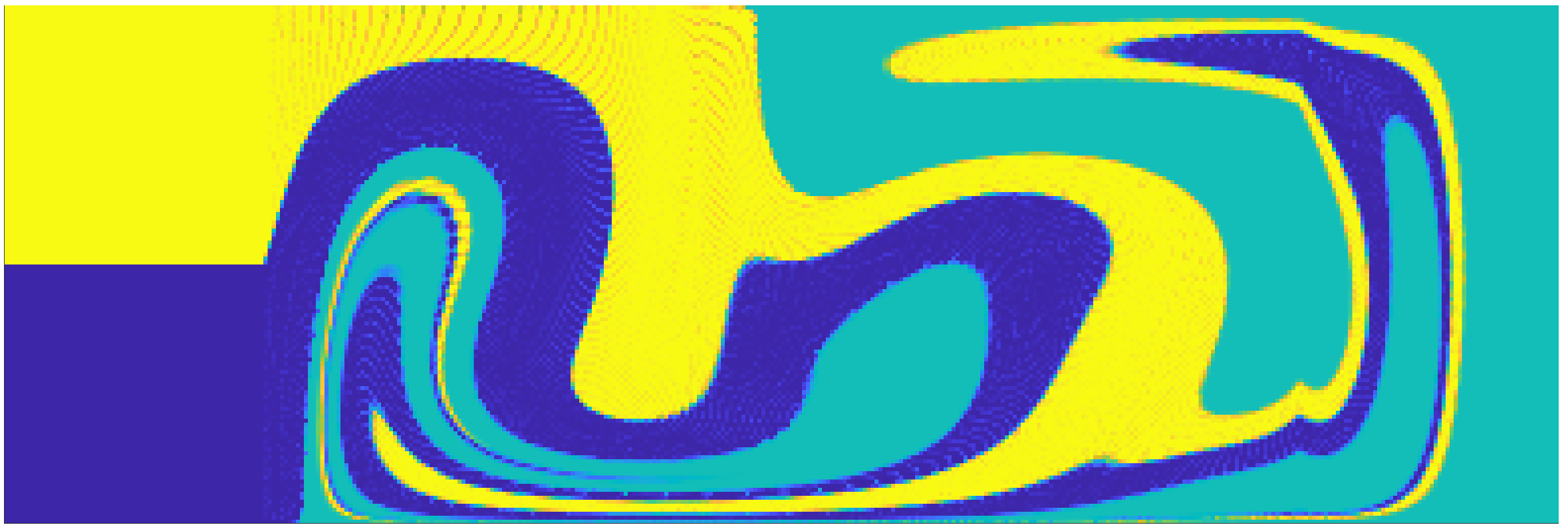} \\
  {\scriptsize $\bm{v}_3$} & {\scriptsize $\bm{v}_4$} \\[2mm]
  \includegraphics[width=0.45\textwidth]{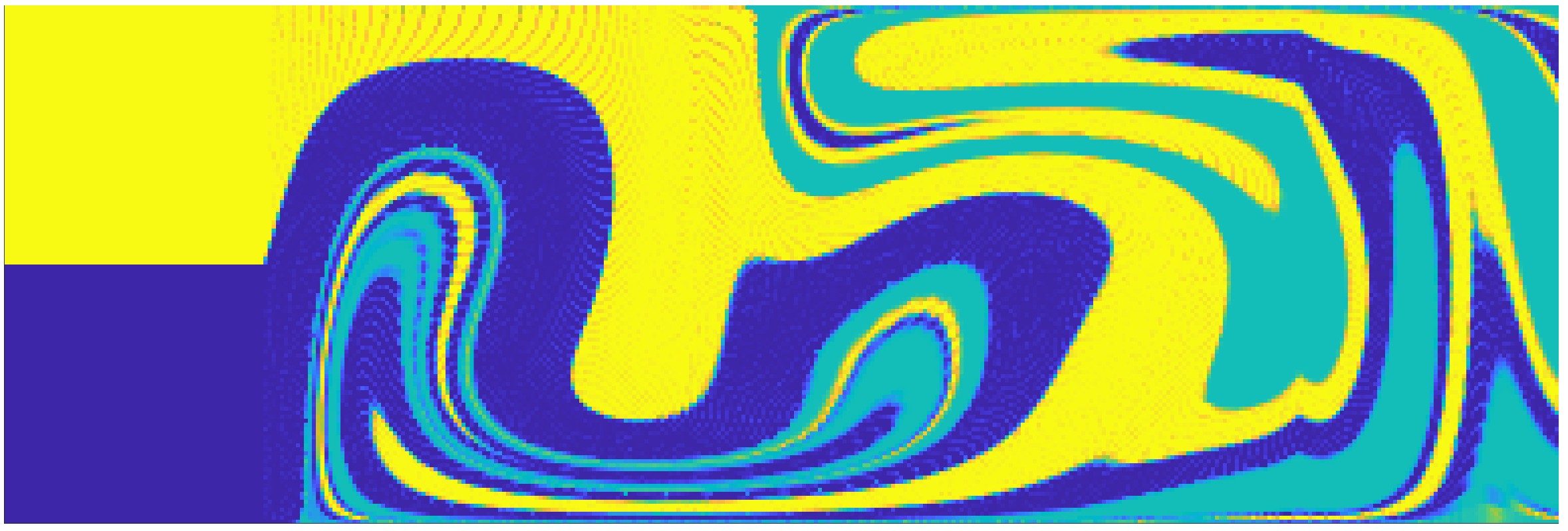} &
  \includegraphics[width=0.45\textwidth]{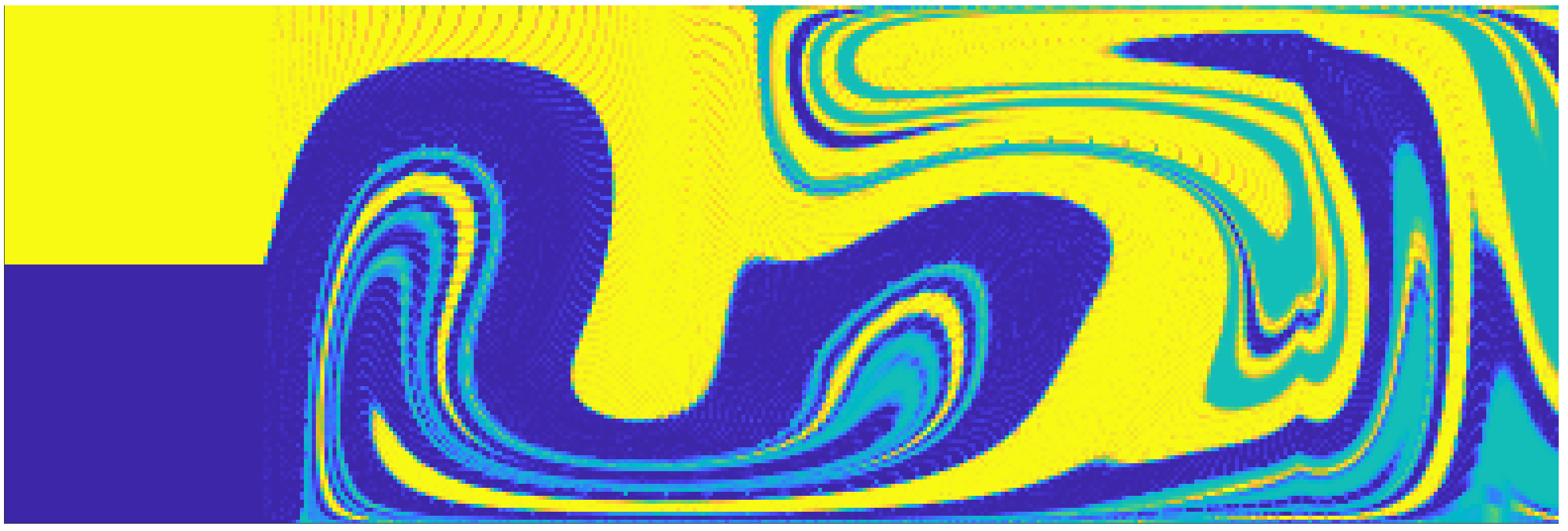} \\
  {\scriptsize $\bm{v}_5$} & {\scriptsize $\bm{v}_6$} \\[2mm]
  \includegraphics[width=0.45\textwidth]{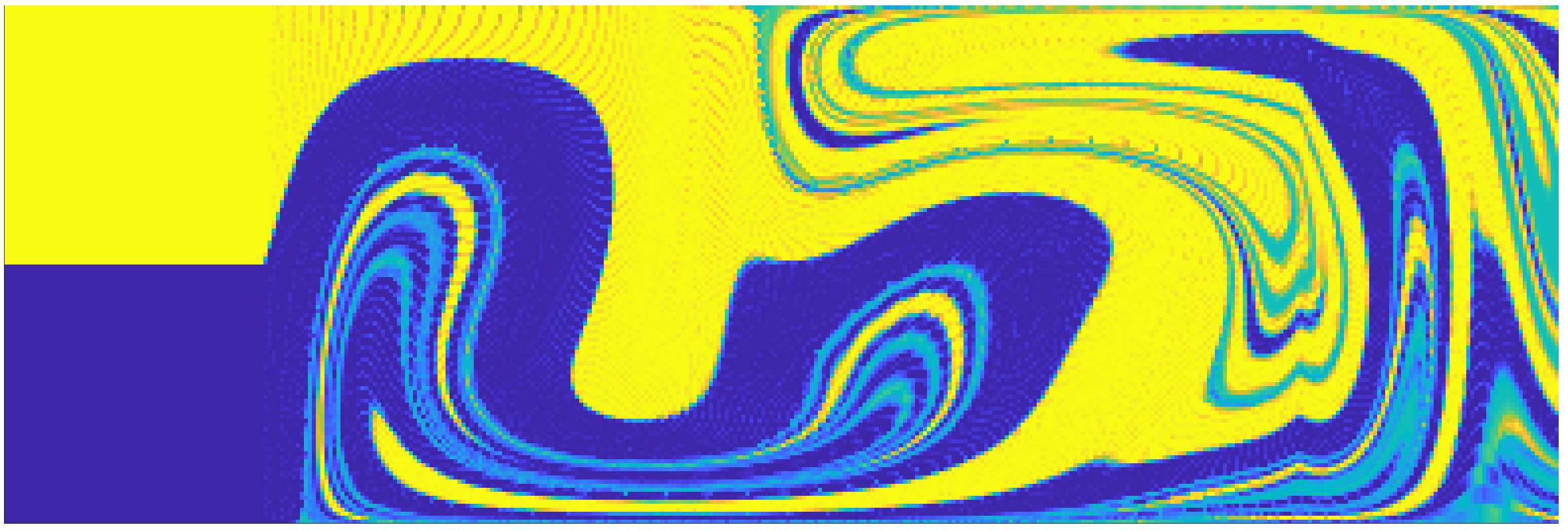} &
  \includegraphics[width=0.45\textwidth]{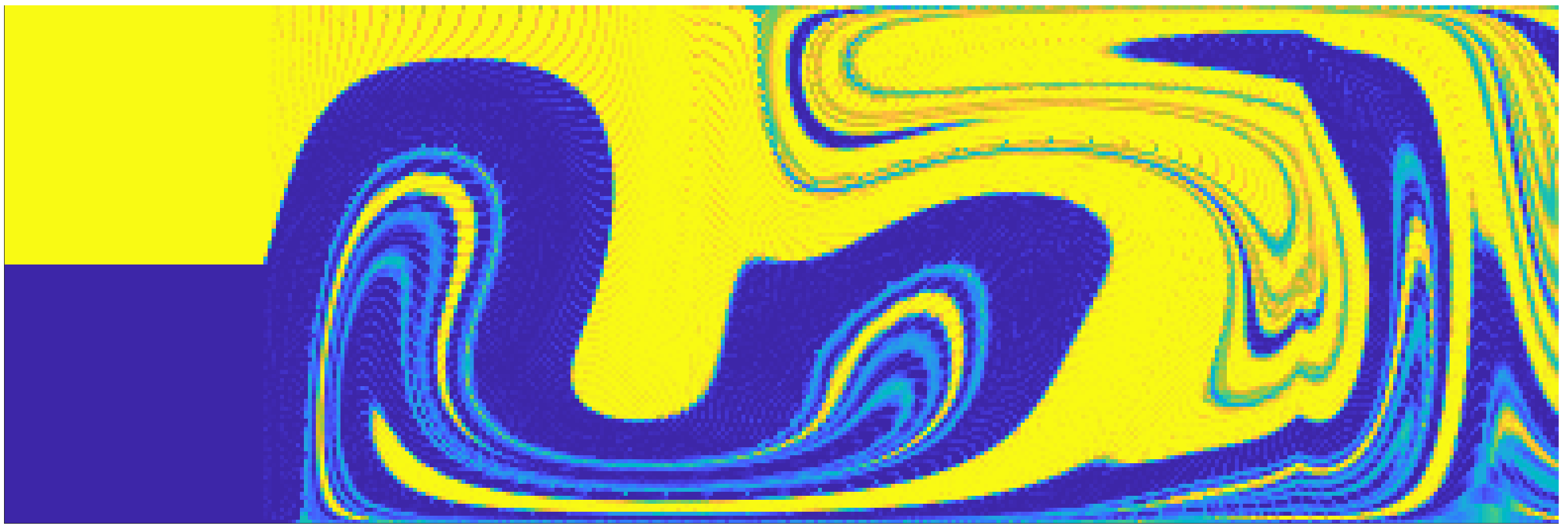} \\
  {\scriptsize $\bm{v}_7$} & {\scriptsize $\bm{v}_8$} \\[2mm]
\end{tabular}
\caption{Evolution of the mass distribution $\bm{v}_k$ for the double gyre mixer with parameters $\alpha=0.5$, $\epsilon=0.4$ and $\beta=0.5$.}\label{fig:dg_vk}
\end{center}
\end{figure}

\paragraph{Evolution of the mass distribution.}
We set the initial mass distribution $\bm{v}_0$ to the zero vector, so that neither of the two fluids is in the mixer at the beginning. The invariant mass distribution is independent of the initial mass distribution, since it leaves the system eventually. For the first experiment, we fix parameters $\alpha=0.5$, $\epsilon=0.4$ and set $\beta=0.5$ for the inflow velocity.
In Fig.~\ref{fig:dg_vk} the first steps of the density evolution $\bm{v}_k$, $k=1, \ldots, 12$, of the set $A$ are illustrated.  With $\bm{v}_0$ being identically zero, the fluids to be mixed are introduced as horizontal stripes into the inlet by adding $\bm{\sigma}$, resulting in $\bm{v}_1=\bm{\sigma}$.  In the next iteration, the inlet contents are mapped into the mixing region and get mixed by $\bm{v}_1 P$. Additionally, new fluid is introduced into the inlet, resulting in $\bm{v}_2=\bm{v}_1 P+\bm{\sigma}$. This is repeated several times. For $k \geq 4$, mixed fluid can be found in the outlet region. For $k\geq 8$ the mixing pattern in the outlet seems to stabilize, indicating that $\bm{v}_k$ quickly converges to the invariant distribution $\bm{v}_{\text{inv}}$. Below, we therefore focus on the invariant distribution $\bm{v}_{\text{inv}}$ and study how well the two fluids are mixed in the outlet region.


\begin{figure}[!htb]
\begin{center}
\includegraphics[width=0.7\textwidth]{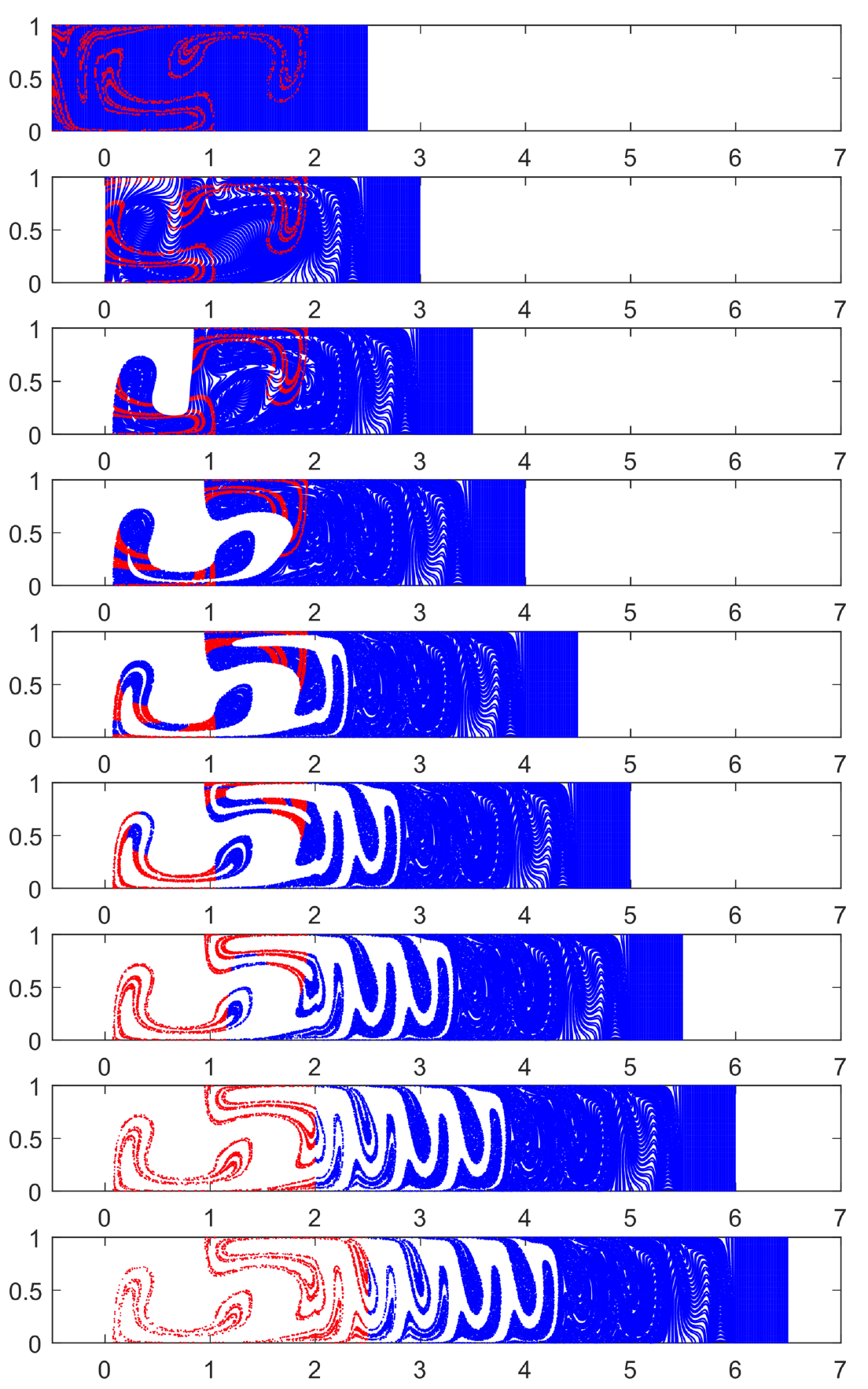}\\
\caption{Evolution of a set of particles for the double gyre mixer on the unbounded domain $X$ with system parameters $\alpha=0.5$, $\epsilon=0.4$ and $\beta=0.5$. Position of the particles after 0, 1, 2, 3, 4, 5, 6, 7 time steps. Particles that stay longer in the mixing region (red) reveal the stable and unstable manifolds.}\label{fig:ink}
\end{center}
\end{figure}

\begin{figure}[!htb]
\begin{center}
\begin{tabular}{c}
 \includegraphics[width=0.6\textwidth]{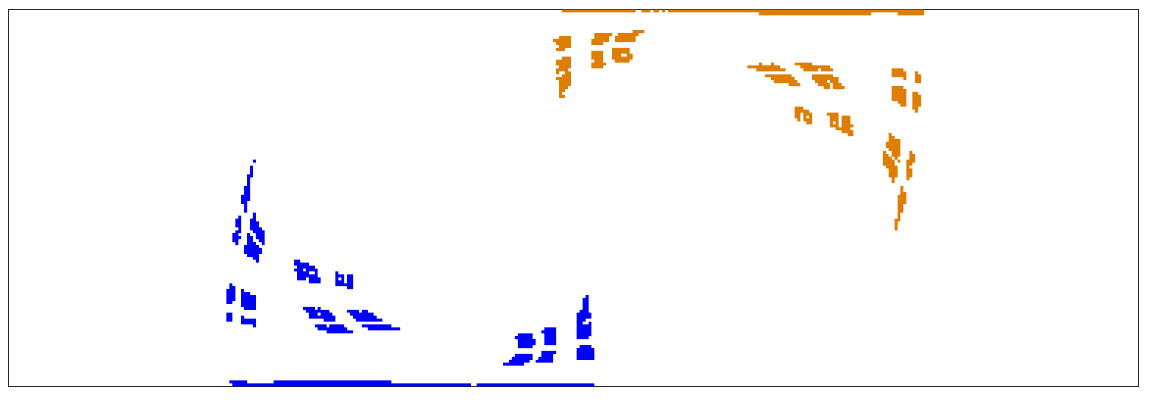} \\[2mm]
\end{tabular}
\caption{Self-communicating classes of $P$ for the double gyre mixer with system parameters $\alpha=0.5$, $\epsilon=0.4$, and $\beta=0.5$. There are two self-communicating classes: Boxes that belong to the first self-communicating class are colored in blue and boxes that belong to the second are colored in orange.}  \label{fig:scc}
\end{center}
\end{figure}

\paragraph{Organizing structures.}
We fix again the parameters $\alpha=0.5$, $\epsilon=0.4$ and $\beta=0.5$. In Fig.~\ref{fig:ink} we follow a set of particles in the unbounded domain $X$ for $8$ time steps. We observe that some particles (red) stay longer in the mixing region. They reveal at time $t=8$ the unstable manifolds of two chaotic saddles. Instead of following particles, we now use the transition matrix $P$ to extract the organizing structures of the system. Using the reachability matrix, we find two self-communicating classes (shown in Fig.~\ref{fig:scc}), which both approximate a chaotic saddle. Neither self-communicating class is accessible from the other, and hence both of them are a maximal class of $P$ and $P^T$.

\begin{figure}[!htb]
\begin{center}
\begin{tabular}{cc}
\includegraphics[scale=0.23]{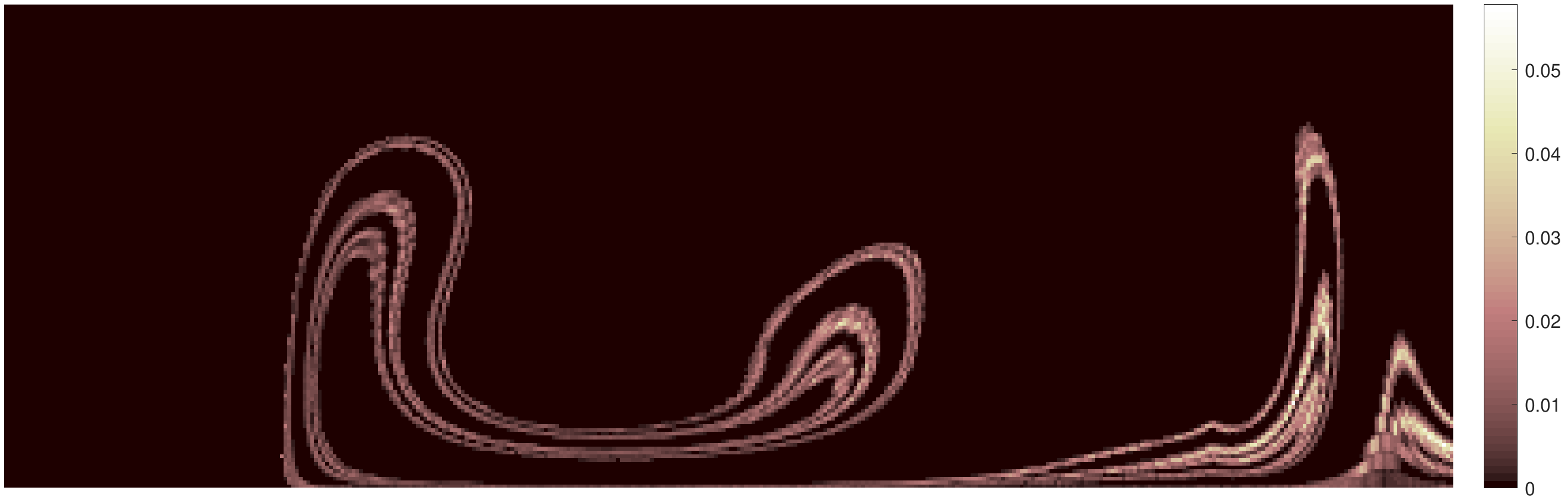} & \includegraphics[scale=0.23]{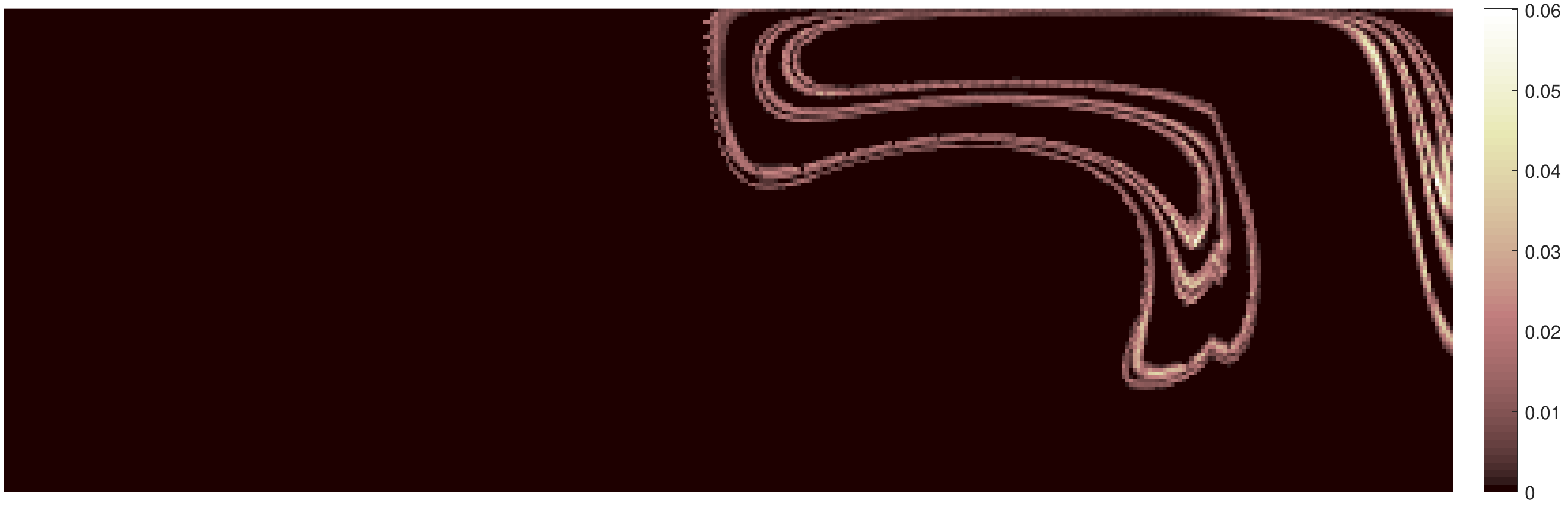} \\
{\scriptsize (a) $\bm{w}_1$ } &  {\scriptsize (d) $\bm{w}_2$ } \\[2mm]
\includegraphics[scale=0.23]{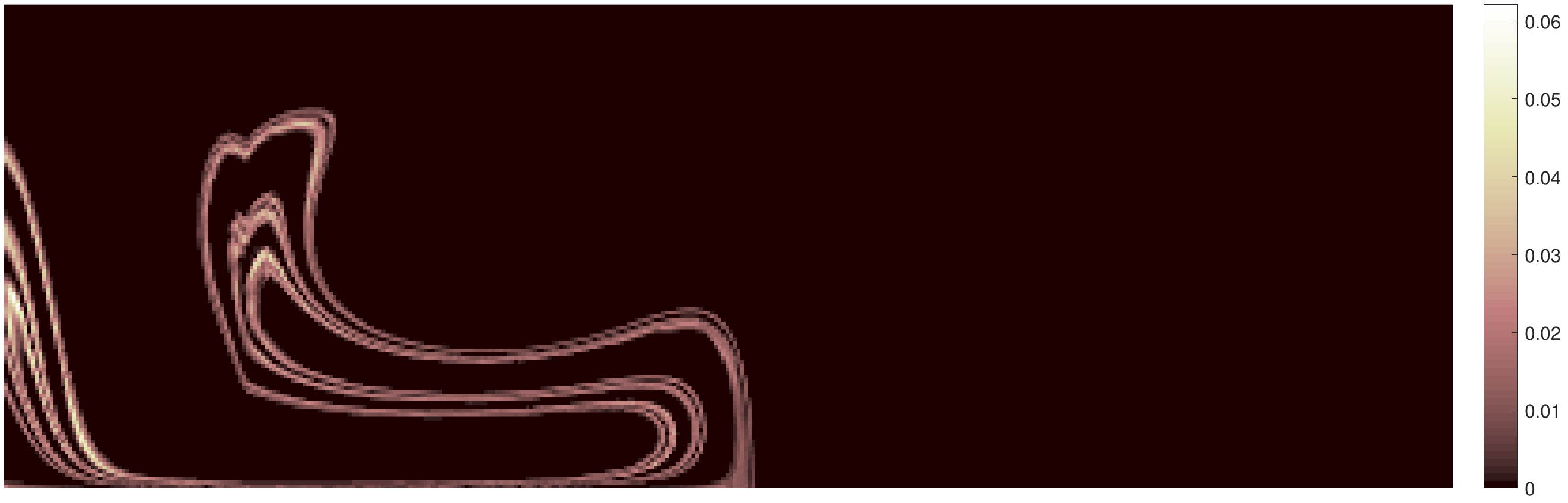} &  \includegraphics[scale=0.23]{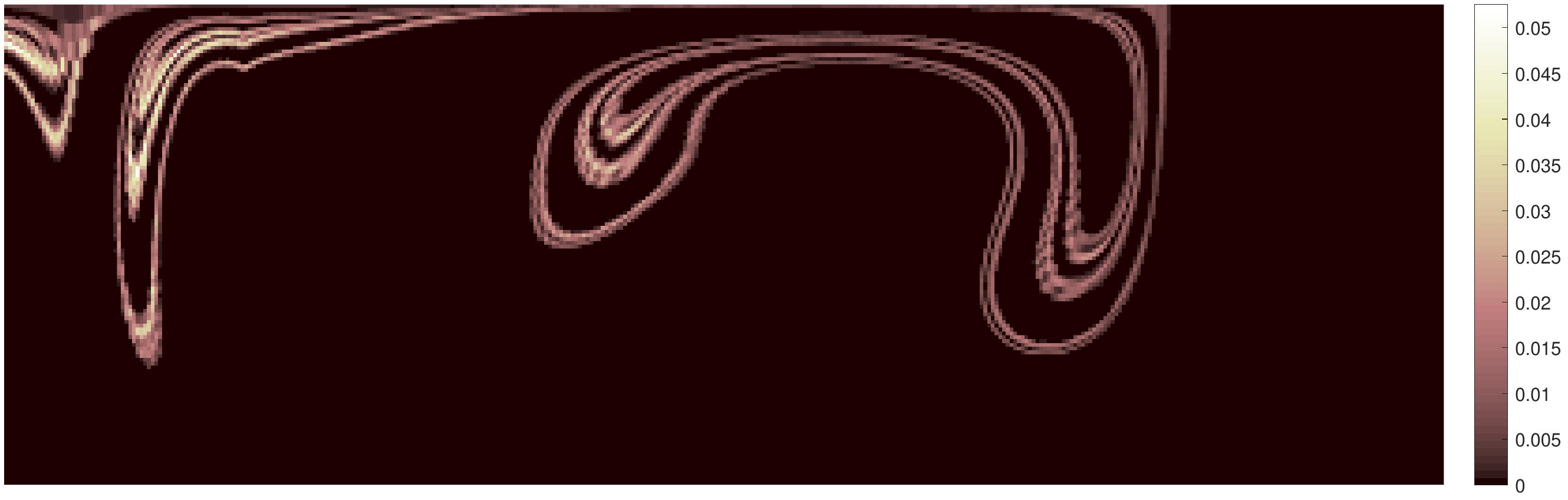} \\
{\scriptsize (b) $\bm{\hat{w}}_1$ } & {\scriptsize (e) $\bm{\hat{w}}_2$ }\\[2mm]
\includegraphics[scale=0.23]{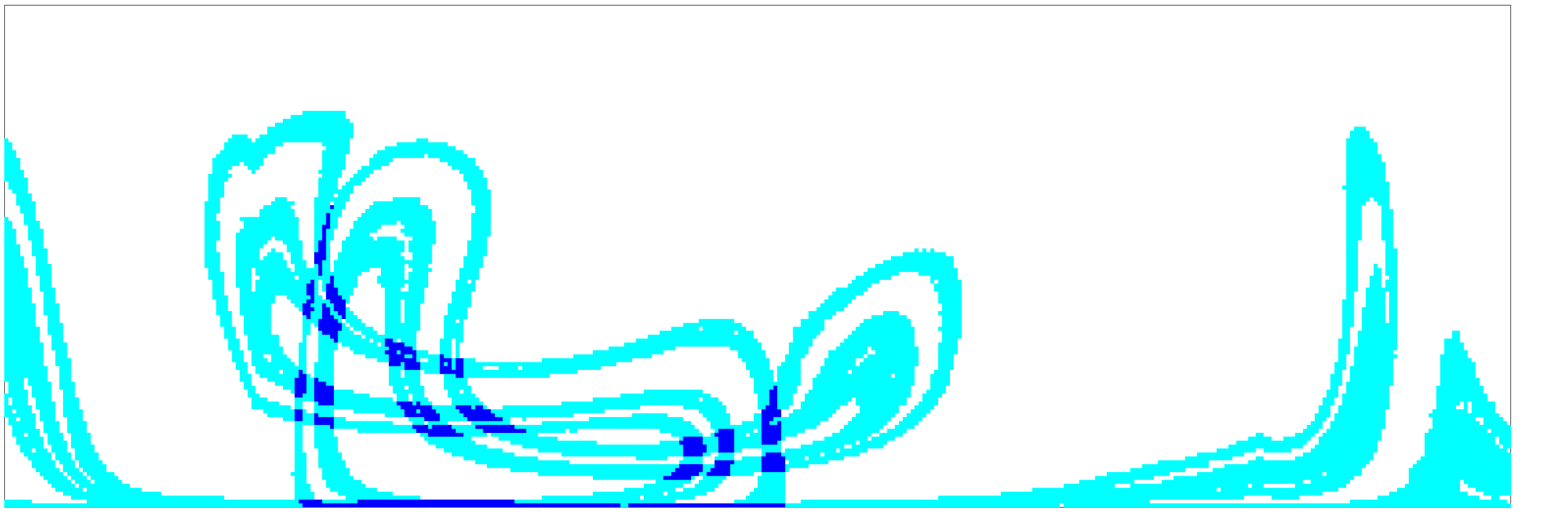} & \includegraphics[scale=0.23]{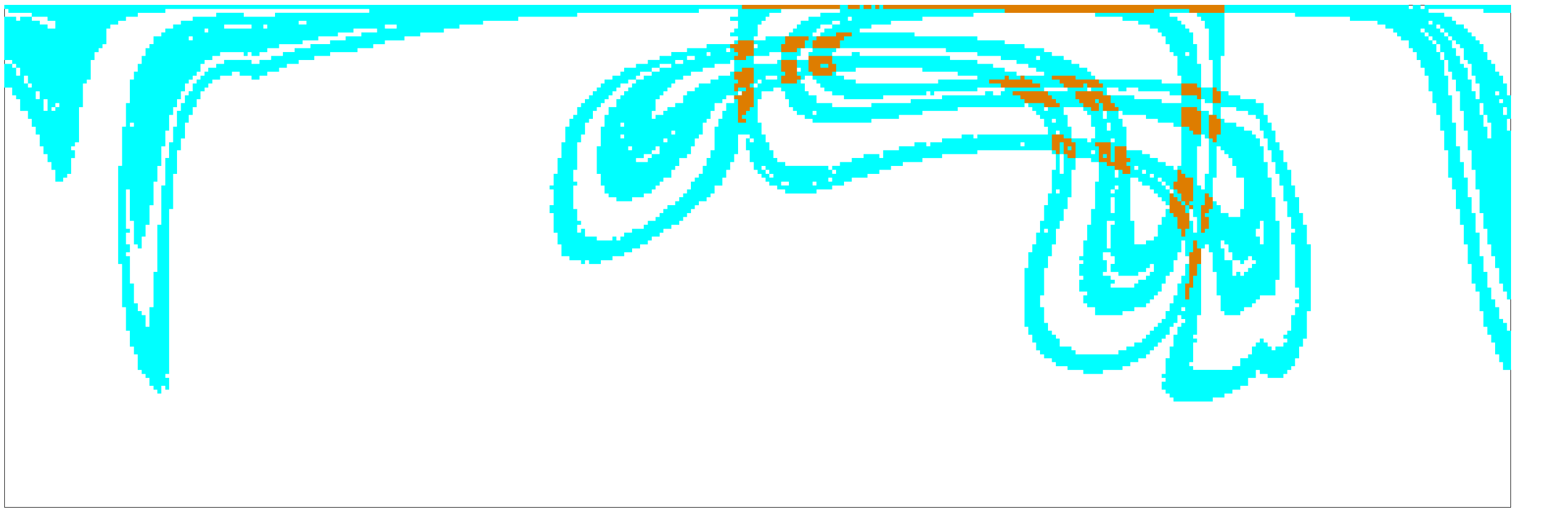}\\
{\scriptsize (c) support of $\bm{w}_1$ and $\bm{\hat{w}}_1$} & {\scriptsize (f) support of $\bm{w}_2$ and $\bm{\hat{w}}_2$} \\[2mm]
 \end{tabular}
\caption{Double gyre mixer with system parameters $\alpha=0.5$, $\epsilon=0.4$, and $\beta=0.5$. First left eigenvector $\bm{w}_1$  (a), first right eigenvector $\bm{\hat{w}}_1$ (b), second left eigenvector $\bm{w}_2$ (d) and second right eigenvector $\bm{\hat{w}}_2$ (e) of $P$. The support of these left and right eigenvectors (c), (f), where entries $<10^{-12}$ are treated as zero. The intersection (dark blue, orange) approximates each a chaotic saddle.} \label{fig:saddle}
\end{center}
\end{figure}

The transition matrix $P$ has the leading real eigenvalues $\lambda_1=0.5817$ and $\lambda_2= 0.5775$. In Fig.~\ref{fig:saddle} the nonnegative left and right eigenvectors for $\lambda_1$ and  $\lambda_2$ are shown. Both the support of the leading first and second left eigenvector each approximate an unstable manifold of the two chaotic saddles. The result matches with the observed structures in Fig.~\ref{fig:ink}.

We can now follow the particles that have stayed longer in the system backward in time. At time $t=0$ they lie close to the stable manifolds.
The supports of the leading first and second right eigenvectors approximate the stable manifolds of the two chaotic saddles. The saddles themselves are approximated by the intersections of the different supports (dark blue and orange).

\begin{figure}[!htb]
\begin{center}
\begin{tabular}{cc}
 \includegraphics[width=0.45\textwidth]{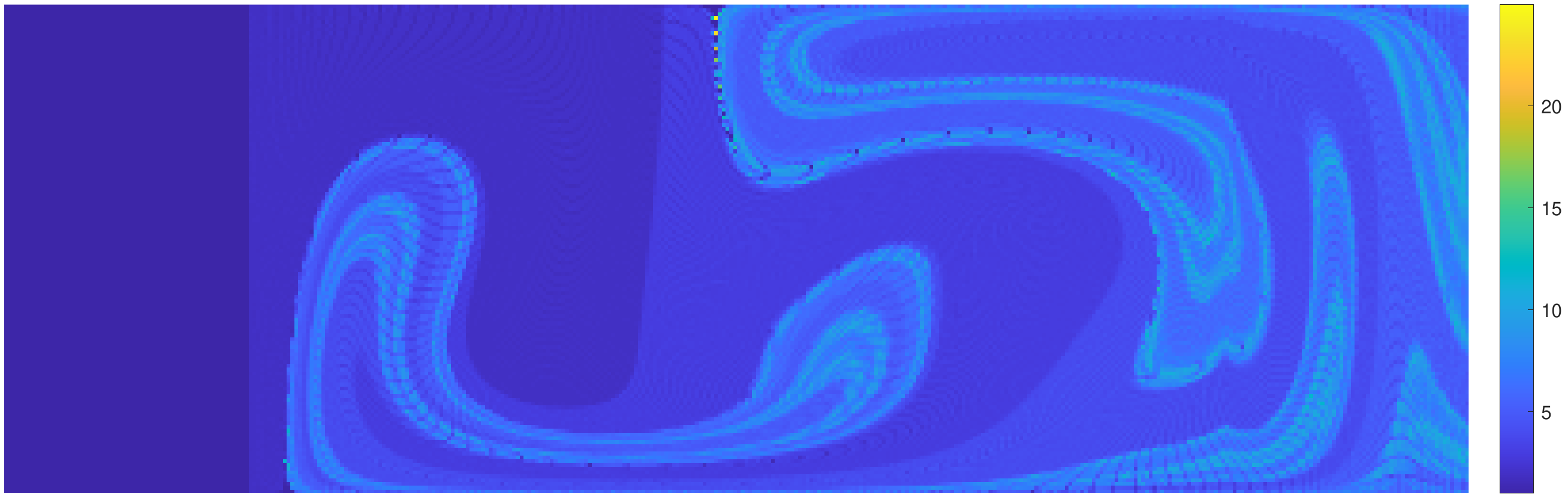} & \includegraphics[width=0.45\textwidth]{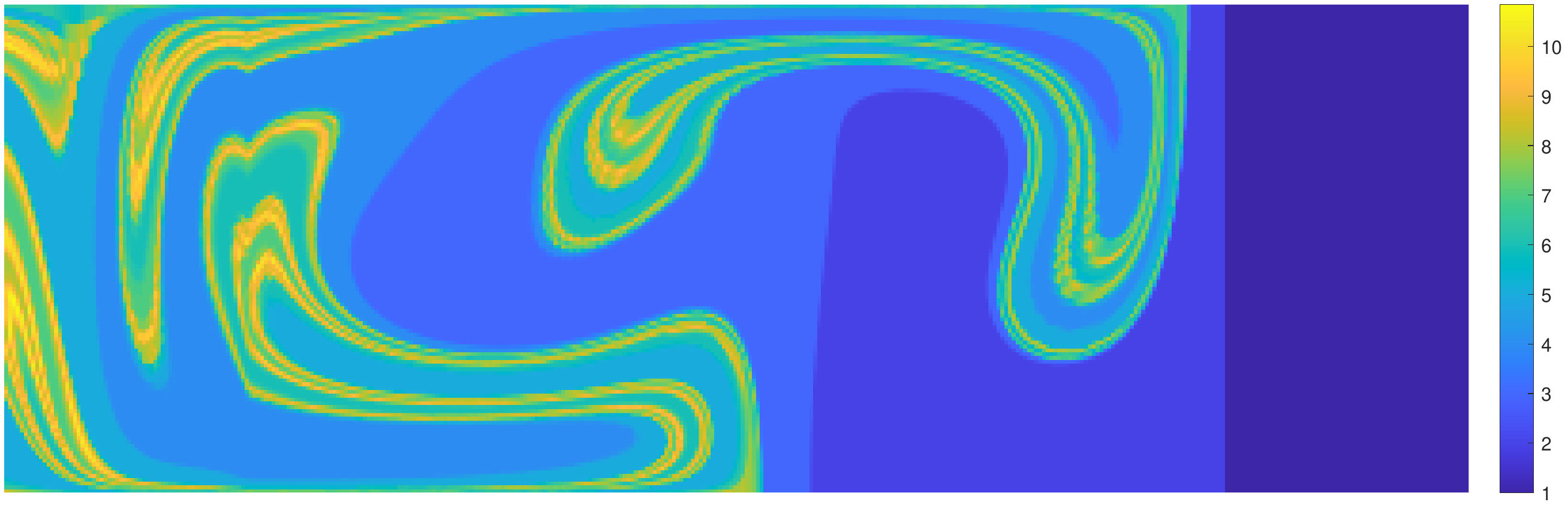}\\[2mm]
\end{tabular}
\caption{Double gyre mixer with system parameters $\alpha=0.5$, $\epsilon=0.4$ and $\beta=0.5$. Expected residence times in backward (left) and forward time (right).} \label{fig:etm}
\end{center}
\end{figure}

The expected residence time for a particle to stay in the domain $A$ of the mixer, given that the particle started in a given box, can be extracted easily from the fundamental matrix. The expected residence times in forward and backward time are shown in Fig.~\ref{fig:etm}. Particles that start in boxes near the stable manifold stay longer in the mixing region. Accordingly, we see that regions of high expected residence times in forward time correspond to regions with high values in the leading right eigenvectors of the transition matrix $P$. Similarly, particles close to the unstable manifold stay longer in the mixing region when we follow them backward in time. Regions of high expected residence times in backward time correspond to regions with high values in the leading left eigenvectors of the transition matrix $P$.

\paragraph{Quantifying mixing and parameter studies.}
For different choices of the parameters $\alpha$, $\epsilon$ and $\beta$, we consider the invariant distribution $\bm{v}_{\text{inv}}$ restricted to the outlet region $A_{3}$ and measure the mixedness of the two fluids.

\begin{figure}[!htb]
\begin{center}
\includegraphics[width=1\textwidth]{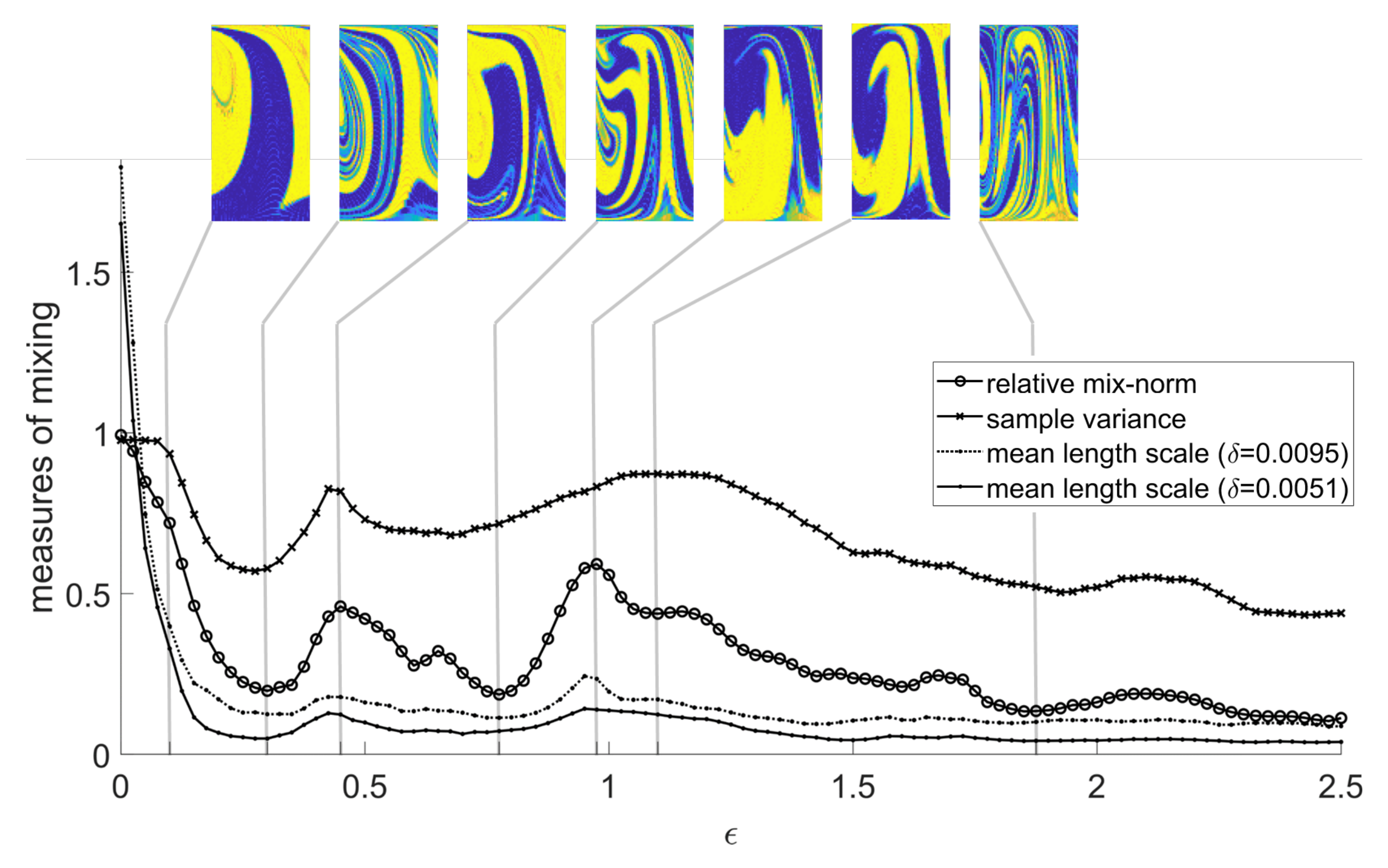}\\
\caption{Different measures of mixing (see legend) applied to $\bm{v}_{\text{inv}}$ restricted to $A_{3}$ for the double gyre mixer for different choices of $\epsilon$. $\alpha=0.5$ and $\beta=0.5$ are fixed. } \label{fig:dg_mm}
\end{center}
\end{figure}

\paragraph{Varying $\epsilon$.} We fix $\alpha=0.5$ and $\beta=0.5$ and vary $\epsilon$ from $0$ to $2.5$. For each setting, we compute the invariant mass distribution $\bm{v}_{\text{inv}}$ as a solution to $(I-P)\bm{v}_{\text{inv}}=\bm{\sigma}$. The different measures of mixing applied to $\bm{v}_{\text{inv}}$ restricted to $A_{3}$ and examples of some invariant mass distributions are shown in Fig.~\ref{fig:dg_mm}. The mixing quality shows an oscillatory behavior instead of a monotonic dependence on $\epsilon$. The different mixing measures quantify this behavior similarly but there are some deviations in the ranking. The two prominent local minima are at $\epsilon=0.275 $ and $\epsilon=0.675 $ for the sample variance and at $\epsilon=0.3$ and $\epsilon=0.775$ for the relative mix-norm. The graph of mean length scale using a bin width tolerance of $\delta=0.0051$ shows similarities to the graph of the sample variance. If one uses a larger bin width tolerance of $\delta=0.0095$, more box pairs are in the neighborhood $N(h)$ for the given distance~$h$. The graph of the calculated mean length scale then mimics the relative mix-norm.

\begin{figure}[!htb]
\begin{center}
\includegraphics[width=1\textwidth]{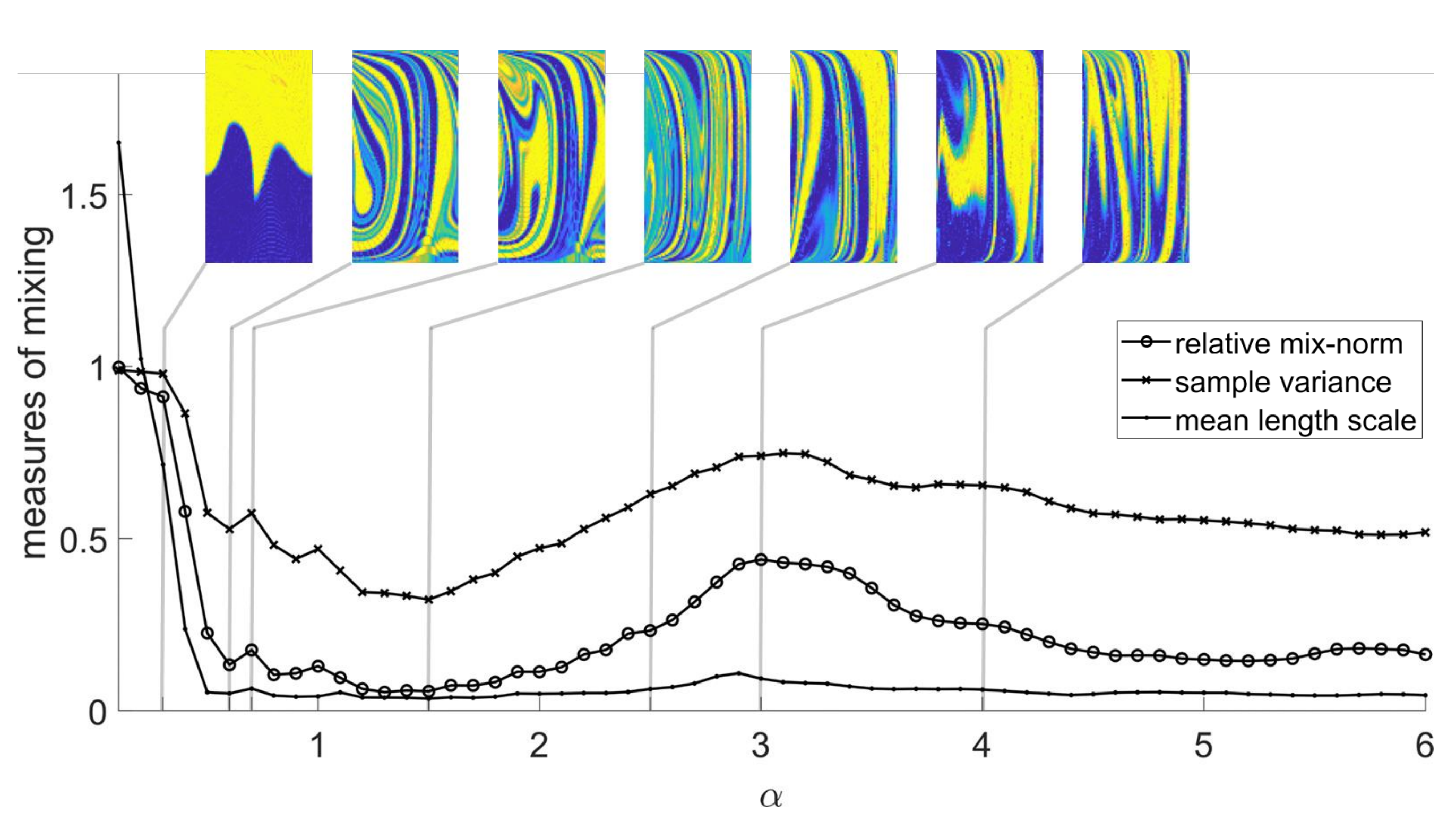}\\
\caption{Different measures of mixing (see legend) applied to $\bm{v}_{\text{inv}}$ restricted to $A_{3}$ for the double gyre mixer for different choices of $\alpha$. $\epsilon=0.25$ and $\beta=0.5$ are fixed.} \label{fig:dg_mm2}
\end{center}
\end{figure}

\paragraph{Varying $\alpha$.} We keep $\beta=0.5$ and fix $\epsilon=0.25$, which was a nearly optimal choice when varying $\epsilon$ for fixed $\alpha$. We have to be sure that there are no dead regions in the mixer and all test particles can eventually leave the system. When $\alpha$ is chosen too large such invariant regions of positive Lebesgue measure can occur. In this case there are extremely high expected residence times and the matrices $I-P$ are almost singular. We therefore restrict the variation of $\alpha$ from $0.1$ to $6$. Examples of invariant mass distributions $\bm{v}_{\text{inv}}$ restricted to $A_{3}$ and the different measures of mixing are shown in Fig.~\ref{fig:dg_mm2}. The curves are again non-monotonic. Optimal mixing occurs for $\alpha=1.5$ and this is captured by all mixing measures. The curve of the mean scale using $\delta=0.0051$ is flatter, and in particular it does not exhibit a pronounced local maximum for $\alpha=3$ (corresponding to poor mixing) as the sample variance and the relative mix-norm do.

\paragraph{Further studies.}
Another parameter that could be changed is the underlying velocity parameter $\beta$, but its impact on mixing is relatively straightforward: If $\beta$ is very large, the influence of the velocity field $\bm{u}_{\mathrm{m}}$ in the mixing region is small and the outlet pattern is poorly mixed, for small $\beta$ the mixing process dominates, resulting in better mixing (not shown).
The mixing results also depend on the source distribution $\bm{\sigma}$~\cite{Thiffeault2008, Gouillart2009, Gouillart2011} and a detailed investigation and optimization is currently underway. In this context, it would be interesting to extend recent transfer-operator-based results on optimal initial tracer patterns in closed flows \cite{Farazmand2017} to the open flow case.

Finally, we note that the suggested numerical framework appears to be very robust; see Appendix \ref{sec:discretize} for a systematic study of the influence of the discretization on the results.

\subsection{The lid-driven cavity mixer}
As a second and more realistic example, we choose the lid-driven cavity flow as mixer \cite{grover2012topological}. This system has a piecewise-steady velocity field, where the flow pattern switches between two states every $\tau/2$ unit of time, where $\tau$ is the period. Its stream function for the period~$t \in [k\tau,(k+1)\tau)$ is
\begin{align*}
\Psi_{\mathrm{m}}(x, y, t)=\begin{cases} U_1 g_1(x,y)+  U_2 g_2(x,y) \text{ for } k \tau \leq t < (k+\frac{1}{2})\tau \\
- U_1 g_1(x,y)+  U_2 g_2(x,y)  \text{ for } (k+\frac{1}{2})\tau \leq t < (k+1)\tau  \end{cases}
\end{align*}
on the domain $[0,a]\times[-b,b]$, where for~$\kappa=1,2$
\begin{equation}
  g_\kappa(x,y)= C_\kappa f_\kappa(y) \sin\left(\frac{\kappa\pi x}{a}\right),
\end{equation}
with
\begin{equation}
  f_\kappa(y)= \frac{2\pi y}{a} \cosh\left(\frac{\kappa\pi b}{ a}\right) \sinh\left(\frac{\kappa\pi y}{ a}\right) - \frac{2\pi b}{a} \sinh\left(\frac{\kappa\pi b}{ a}\right) \cosh\left(\frac{\kappa\pi y}{ a}\right)
\end{equation}
and
\begin{equation}
  C_\kappa= \frac{a^2}{2 \kappa \pi^2 b} \left[\frac{a}{2 \kappa \pi b} \sinh\left(\frac{2 \kappa \pi b}{a}\right)+1\right]^{-1}.
\end{equation}
The ratio ${U_2}/{U_1}$ regulates the streamline pattern of the flow.

\begin{figure}[!htb]
\begin{center}
\includegraphics[width=0.45\textwidth]{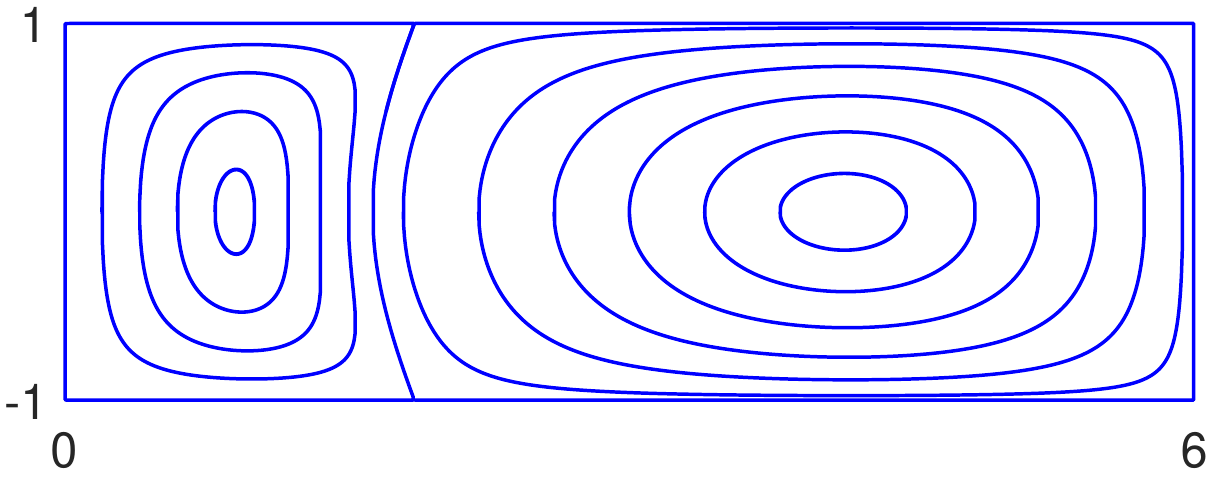}\hspace{1em}   \includegraphics[width=0.45\textwidth]{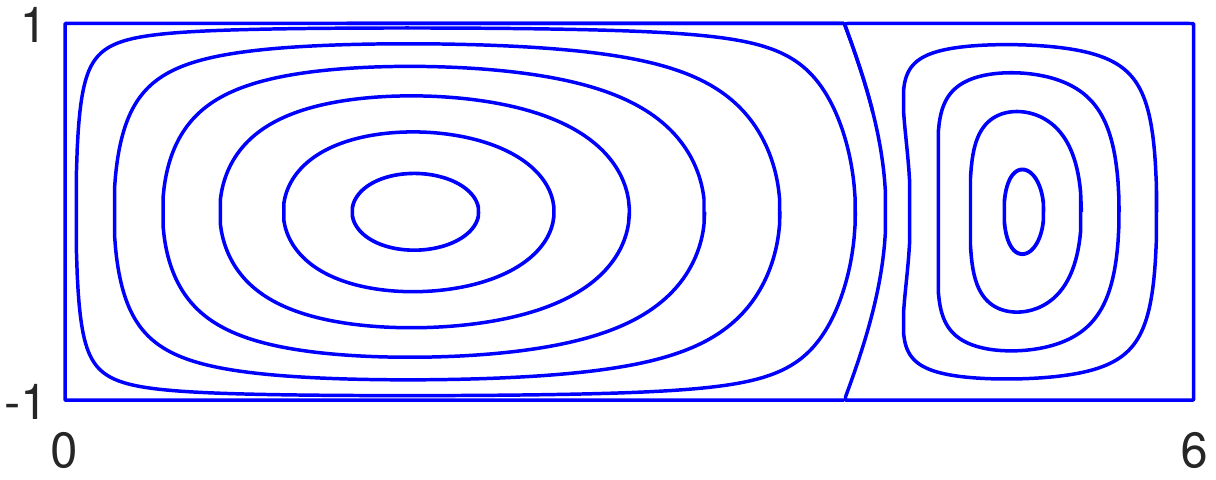}\\
\caption{Streamlines of the lid-driven cavity flow with $U_1=9$ and $U_2=8$ at $t=0$ (left) and $t=0.5$ (right).}\label{fig:liddriven}
\end{center}
\end{figure}

We fix parameters $\tau=1$, $a=6$ and $b=1$. The streamlines $\bm{u}_{\mathrm{m}}$ on the mixing region $X_2=A_{2}=[0,6]\times[-1,1]$ with $U_1=9$ and $U_2=8$ for time instances $t = 0$ and $t = 0.5$ are shown in Fig.~\ref{fig:liddriven}. For the background flow, we choose here $\Psi_{\mathrm{b}}= \beta y$, with~$\beta =1$. We divide the domain $A=[-1,7]\times [-1,1]$ in $2^{16}$ square boxes of side length $0.0156$. The boxes in $A_{1}=[-1,0]\times[-1,1]$ correspond to the source states. We initialize 100 test particles on a regular grid in each box and integrate with the classical Runge-Kutta method with step size $h=0.01$ from $t=0$ to $1$.

\begin{figure}[!htb]
\begin{center}
\begin{tabular}{cc}
  \includegraphics[width=0.45\textwidth]{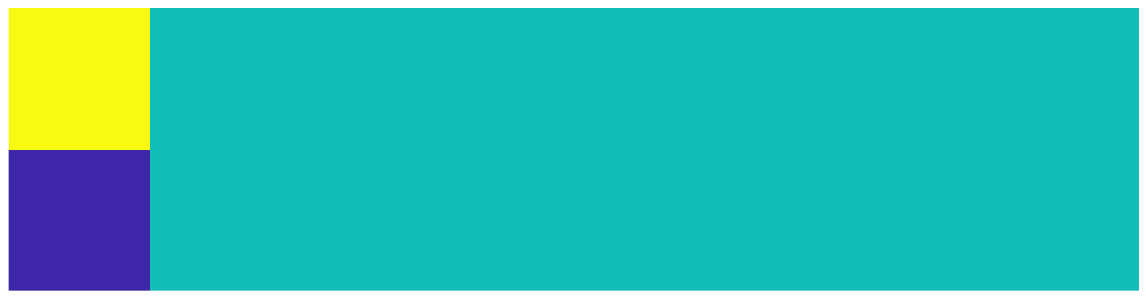} &
  \includegraphics[width=0.45\textwidth]{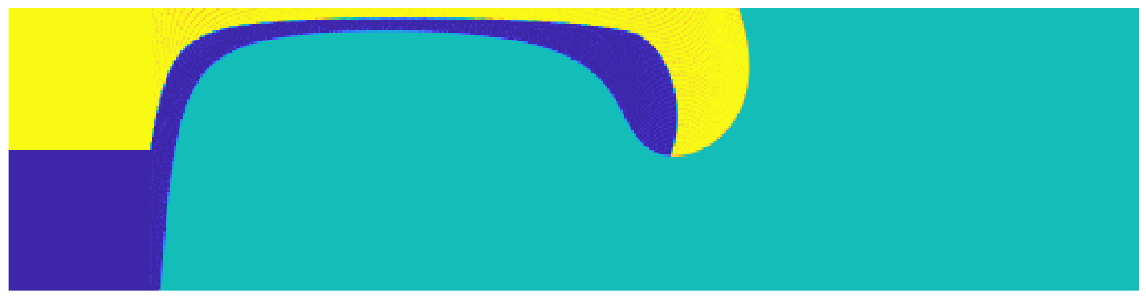} \\
  {\scriptsize $\bm{v}_1$} & {\scriptsize $\bm{v}_2$} \\[2mm]
  \includegraphics[width=0.45\textwidth]{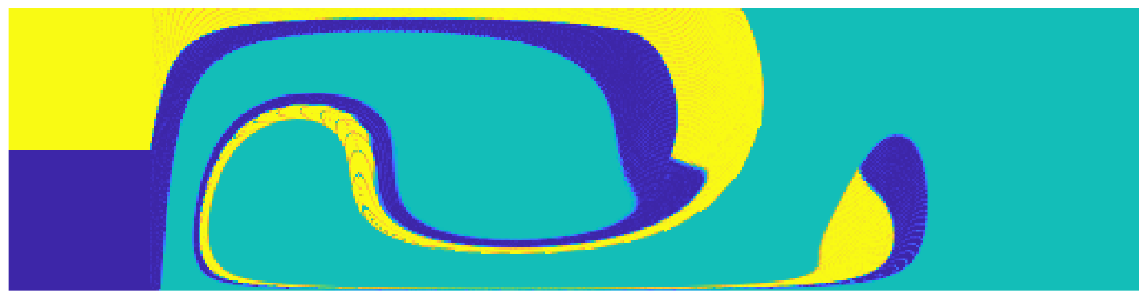} &
  \includegraphics[width=0.45\textwidth]{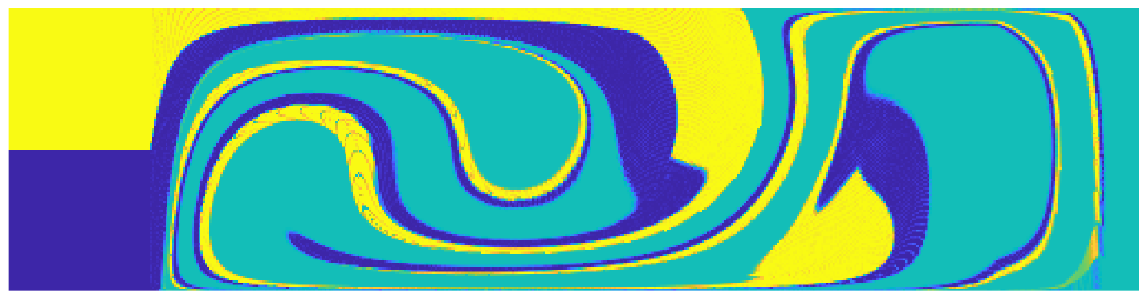} \\
  {\scriptsize $\bm{v}_3$} & {\scriptsize $\bm{v}_4$} \\[2mm]
  \includegraphics[width=0.45\textwidth]{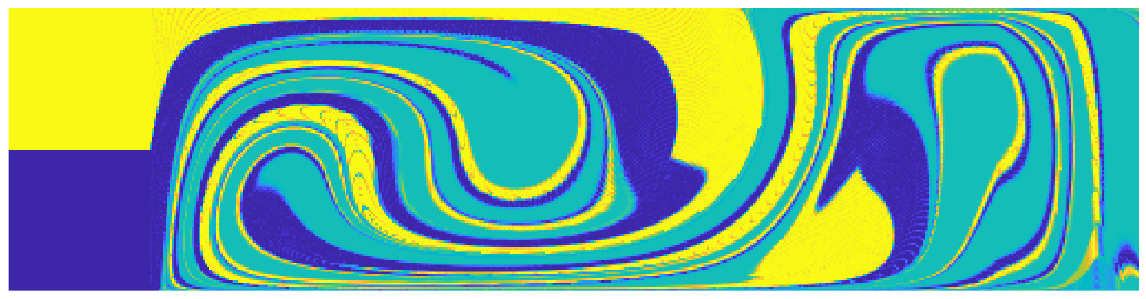} &
  \includegraphics[width=0.45\textwidth]{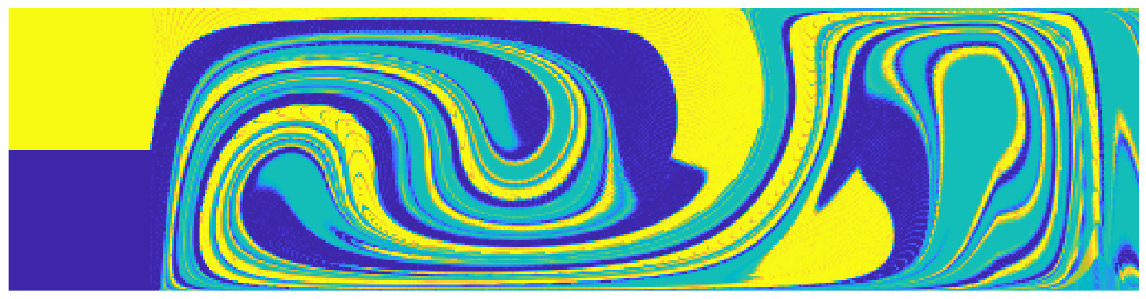} \\
  {\scriptsize $\bm{v}_5$} & {\scriptsize $\bm{v}_6$} \\[2mm]
  \includegraphics[width=0.45\textwidth]{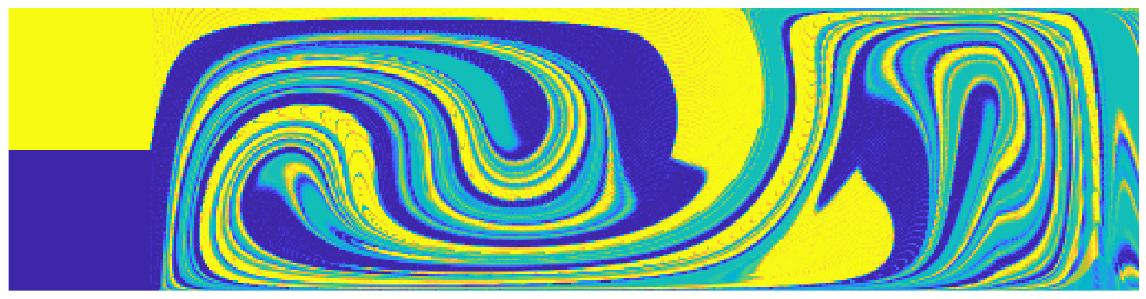} &
  \includegraphics[width=0.45\textwidth]{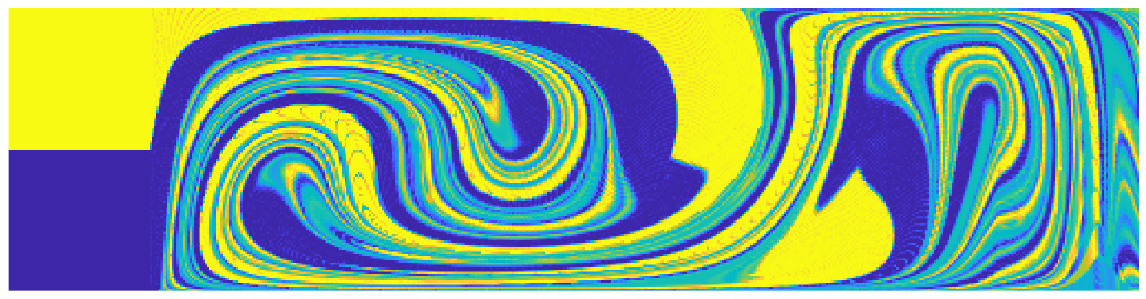} \\
  {\scriptsize $\bm{v}_7$} & {\scriptsize $\bm{v}_8$} \\[2mm]
\end{tabular}
\caption{Evolution of the mass distribution $\bm{v}_k, k=1,2,3, \ldots, 12$, for the lid-driven cavity mixer with system parameters  $U_1=9$, $U_2=8$ and $\beta=1$.}\label{fig:lm_vk}
\end{center}
\end{figure}

\paragraph{Evolution of the mass distribution.}
We set the initial mass distribution $\bm{v}_0$ to $0$, and fix the parameters $U_1=9$  and $U_2=8$ and $\beta=1$. In Fig.~\ref{fig:lm_vk} the first steps of the density evolution $\bm{v}_k$, $k=1, \ldots, 8$, on the set $A$ are illustrated. At each iteration the source distribution $\bm{\sigma}$ is added, which introduces the fluids to be mixed as horizontal stripes into the inlet.

\begin{figure}[!htb]
\begin{center}
\begin{tabular}{c}
\includegraphics[width=0.7\textwidth]{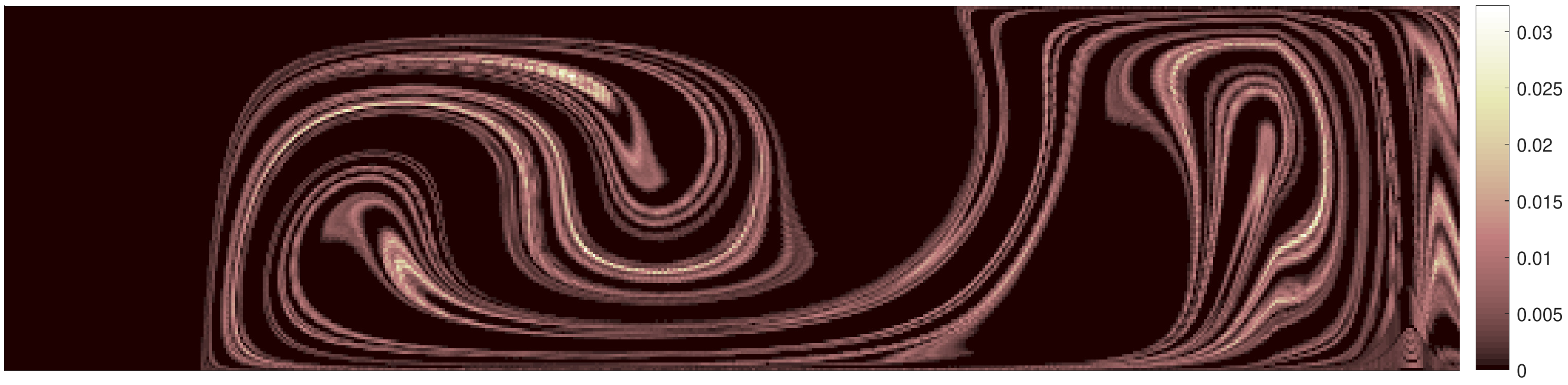} \\
{\scriptsize (a) $\bm{w}_1$ }\\
 \includegraphics[width=0.7\textwidth]{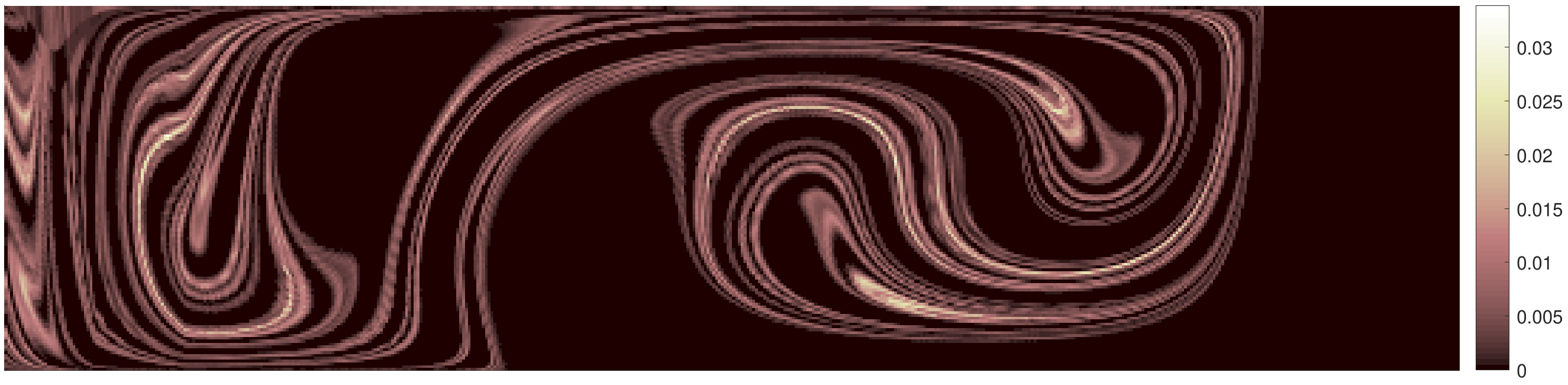} \\
{\scriptsize (b) $\bm{\hat{w}}_1$ }\\[2mm]
 \includegraphics[width=0.7\textwidth]{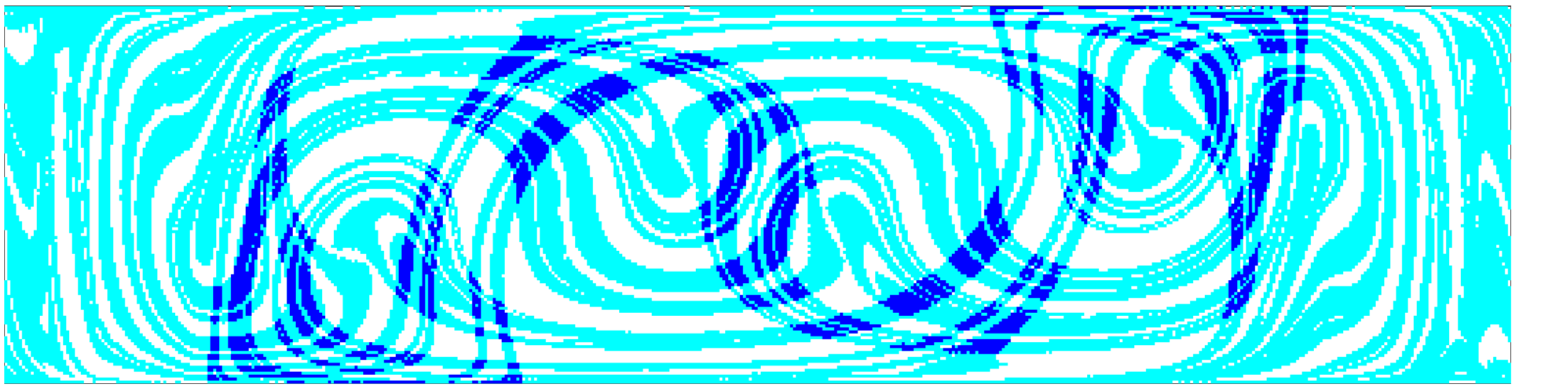}\\
{\scriptsize (c) support of $\bm{w}_1$ and $\bm{\hat{w}}_1$} \\[2mm]
\end{tabular}
\caption{Lid-driven cavity mixer with system parameters $U_1=9$, $U_2=8$ and $\beta=1$. First left eigenvector $\bm{w}_1$ (a), first right eigenvector $\bm{\hat{w}}_1$ (b) and the support of these eigenvectors, where entries $<10^{-3}$ are treated as zero. The intersection (dark blue) approximates the chaotic saddle.}\label{fig:lidsaddle}
\end{center}
\end{figure}

\paragraph{Organizing structures.}
We fix again the parameters $U_1=9$ and $U_2=8$ and $\beta=1$. The transition matrix $P$ has the leading real eigenvalue $\lambda_1= 0.7709$. In Fig.~\ref{fig:lidsaddle} the left and right eigenvector for $\lambda_1$ are shown. The intersection of the support of these eigenvectors (dark blue) reveals the chaotic saddle of the system.


\begin{figure}[!htb]
\begin{center}
\includegraphics[width=1\textwidth]{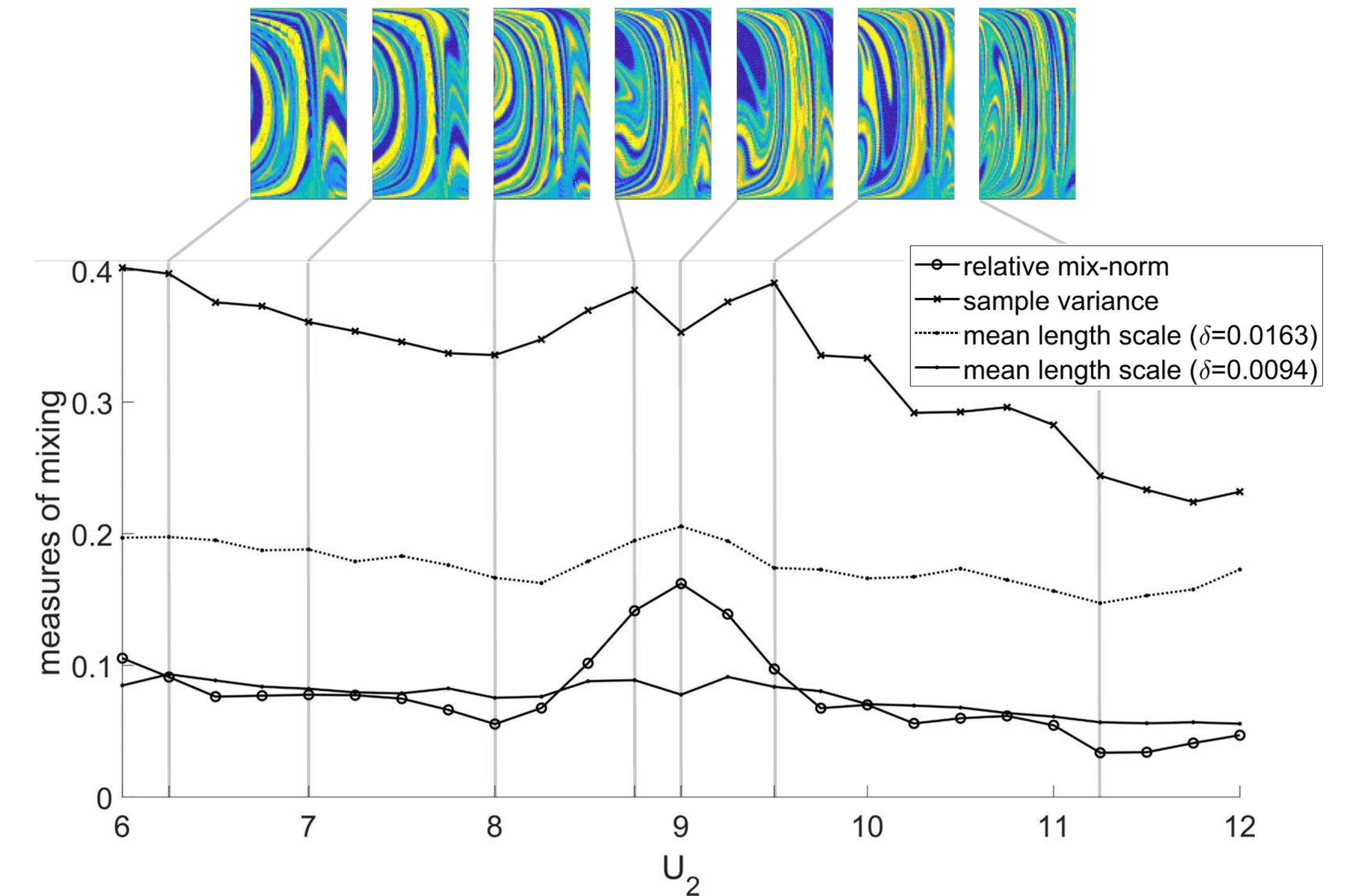}
\caption{Different measures of mixing (see legend) applied to $\bm{v}_{\text{inv}}$ restricted to $A_{3}$ for the lid-driven cavity mixer for different choices of $U_2$. $U_1=9$ and $\beta=1$ are fixed.}  \label{fig:lm_mm2}
\end{center}
\end{figure}

\paragraph{Quantifying mixing and parameter studies.}
We fix $U_1=9$ and vary the parameter $U_2$. We consider $U_2= 6, 6.25, \ldots,12$. Examples of invariant mass distributions restricted to $A_{3}$ and the different measures of mixing are shown in Fig.~\ref{fig:lm_mm2}. For this range of parameters the mixing patterns are all relatively well mixed. Around $U_2=9$ one observes patterns that are less mixed. The graphs of the sample variance and the relative mix-norm show here a difference in the quantification of the mixing quality. The relative mix-norm has one local maximum at $U_2=9$ whereas the sample variance has two local maxima at $U_2=8.75$ and $U_2=9.5$.

The graph of the mean length scale using a bin width tolerance $\delta=0.0094$ shows also two local maxima, at $U_2=8.75$ and $U_2=9.25$. If one uses the larger bin width tolerance $\delta=0.0163$, as in the graph of the relative mix-norm, the graph of the calculated mean length scale then has one local maximum at $U_2=9$.

\section{Conclusion and outlook}
\label{sec:discussion}

In this work, we applied a transfer operator approach for studying mixing processes in open-flow systems. We conducted parameter studies in two sample systems, the double gyre mixer and the more realistic lid-driven cavity mixer. For the quantification of finite-time mixing of two different fluids, we applied several frequently used mixing measures, such as sample variance, mean length scale, and a multiscale mix-norm. While the different measures give consistent results, the mix-norm appears to be most robust with respect to numerical parameters (e.g. box size, number of test particles). In open systems, the chaotic saddle and its stable and unstable manifolds are known to organize the mixing processes. We demonstrate that the leading left and right eigenvectors or the self-communicating classes of the transition matrix can be used to approximate these structures. However, a direct relation between spectral properties of the substochastic transition matrix and the finite-time mixing properties observed is not obvious, as there is no information of the source distribution in the transition matrix $P$. We will explore spectral mixing measures in future work.

The mixing results depend both on the underlying dynamics encoded in the transition matrix and on the source distribution $\bm{\sigma}$. In this paper, we have restricted our study to changes in mixing quality under parameter variations of the flow. A detailed investigation of the  influence of the source distribution $\bm{\sigma}$ is currently underway. It would be interesting to extend recent transfer-operator based results on optimal initial tracer patterns in closed flows \cite{Farazmand2017} to open flows and to combine the open-flow transfer operator framework with optimization schemes to maximize mixing, such as in \cite{froyland2016optimal}.

While the computational studies presented in this paper are restricted to two-dimensional examples with a constant background flow, the extension of our approach to the three-dimensional setting is straightforward, as is the consideration of more realistic background flows (e.g. parabolic flow). The study of nonautonomous, aperiodic flows is more challenging. In practice, it requires the computation and application of  time-dependent transition matrices $P_i$ on finite time intervals $[t_{i-1}, t_{i}]$ of a fixed length $\tau=t_i-t_{i-1}$,  $i>0$.  Given $\bm{v}_0$ and source distribution $\bm{\sigma}$, one then considers $\bm{v}_i=\bm{v}_{i-1}P_i + \bm{\sigma}$. The (now time-dependent) mixing patterns obtained from restricting $\bm{v}_i$ to the outflow region can be studied using the different mixing measures. We will address this in future work.
For aperiodic flows it might be promising to consider the recent data-based approaches by which one can define a random walk between single trajectories \cite{Banisch2017,Padberg2017,Banisch2019}.  Finally, the study of open flows from velocity fields that are only available on part of the domain (such as patches of the global ocean) poses another methodological challenge, as already investigated for the approximation of transport barriers \cite{Tang2010}.

\section*{Acknowledgments}

This research has been supported by the Deutsche Forschungsgemeinschaft within the Priority Programme DFG-SPP 1881 on Turbulent Superstructures. AK and KPG thank Sanjeeva Balasuriya for fruitful discussions.

\bibliography{biblio}

\appendix

%
%
%
%
%
%
%
%

\section{Superposition of stream functions}
\label{sec:bump}

Let $\Psi_{\mathrm{b}}$ be the stream function of the constant background flow, which shifts fluid through the mixing region and acts on $X$, and $\Psi_{\mathrm{m}}$ the stream function of the time-periodic mixer, which acts only on $X_2$.  A smooth superposition of the two stream functions $\Psi_{\mathrm{b}}$ and $\Psi_{\mathrm{m}}$ has the following form:
\begin{equation}
  \Psi(x,y,t)=\Psi_{\mathrm{b}}(x,y)+B_{X_2}(x,y)\Psi_{\mathrm{m}}(x,y,t),
\end{equation}
where $B_{X_2}:\mathbb{R}^2 \to [0, 1]$ is a standard bump function supported on $X_2$. Thus, it is zero outside $X_2$ and sharply increases to one in $X_2$.  We set $\hat{\Psi}_{\mathrm{m}}(x,y,t):=B_{X_2}(x,y)\Psi_{\mathrm{m}}(x,y,t)$ and thus write $\Psi=\Psi_{\mathrm{b}}+\hat{\Psi}_{\mathrm{m}}$.
The smooth velocity field on $X$ is then obtained as:
\begin{equation*}
  \dot{x} =  \frac{\partial \Psi_{\mathrm{b}}}{\partial y} + \frac{\partial \hat{\Psi}_{\mathrm{m}}}{\partial y},\qquad
\dot{y} = - \frac{\partial \Psi_{\mathrm{b}}}{\partial x} - \frac{\partial \hat{\Psi}_{\mathrm{m}}}{\partial x}.
\end{equation*}
Written differently, the velocity field on $X$ has the form
\begin{equation}
  \bm{u}(x,y,t)=\begin{cases} \bm{u}_{\mathrm{b}}(x,y),  & \text{for } (x,y) \in X_1 \cup X_3\\  \bm{u}_{\mathrm{b}}(x,y) + \hat{\bm{u}}_{\mathrm{m}}(x,y,t), &\text{for } (x,y)\in X_2\end{cases}
\end{equation}
where $\bm{u}_{\mathrm{b}}$ is a constant homogeneous velocity field and $\hat{\bm{u}}_{\mathrm{m}}$ is the velocity field of the time-periodic mixer.

In our first example, the double gyre system, we can use the bump function $B(x):=B_{X_2}(x,y)=\exp(\frac{1}{\sigma^2}(\frac{2}{x(x-2)}+1))\bm{1}_{(0,2)}(x)$, which only depends on the $x$-coordinate, to define the stream function $\hat{\Psi}_{\mathrm{m}}(x,y,t) = B(x) \Psi_{\mathrm{m}}(x,y,t)$ and finally
\begin{equation}
  \Psi(x,y,t)=\Psi_{\mathrm{b}}(x,y)+\hat{\Psi}_{\mathrm{m}}(x,y,t).
\end{equation}
For large $\sigma^2$ (e.g. $\sigma^2=100$) $\hat{\Psi}_{\mathrm{m}}(x,y,t)$ and ${\Psi}_{\mathrm{m}}(x,y,t)$ are nearly indistinguishable, so that numerically it suffices to consider the discontinuous velocity field
\begin{equation}
  \bm{u}(x,y,t)=\bm{u}_{\mathrm{b}}(x,y)+\bm{u}_{\mathrm{m}}(x,y,t) \bm{1}_{[0,2]}(x),
\end{equation}
where $\bm{u}_{\mathrm{m}}(x,y,t)$ is derived from the stream function $\Psi_{\mathrm{m}}$.

Analogously, for our second example, the lid-driven cavity flow, the bump function $B(x):=B_{X_2}(x,y)=\exp(\frac{1}{\sigma^2}(\frac{9}{x(x-3)}+1))\bm{1}_{(0,6)}(x)$ can be used.

\section{Influence of discretization}
\label{sec:discretize}
\begin{figure}
\begin{center}
 \includegraphics[width=0.8\textwidth]{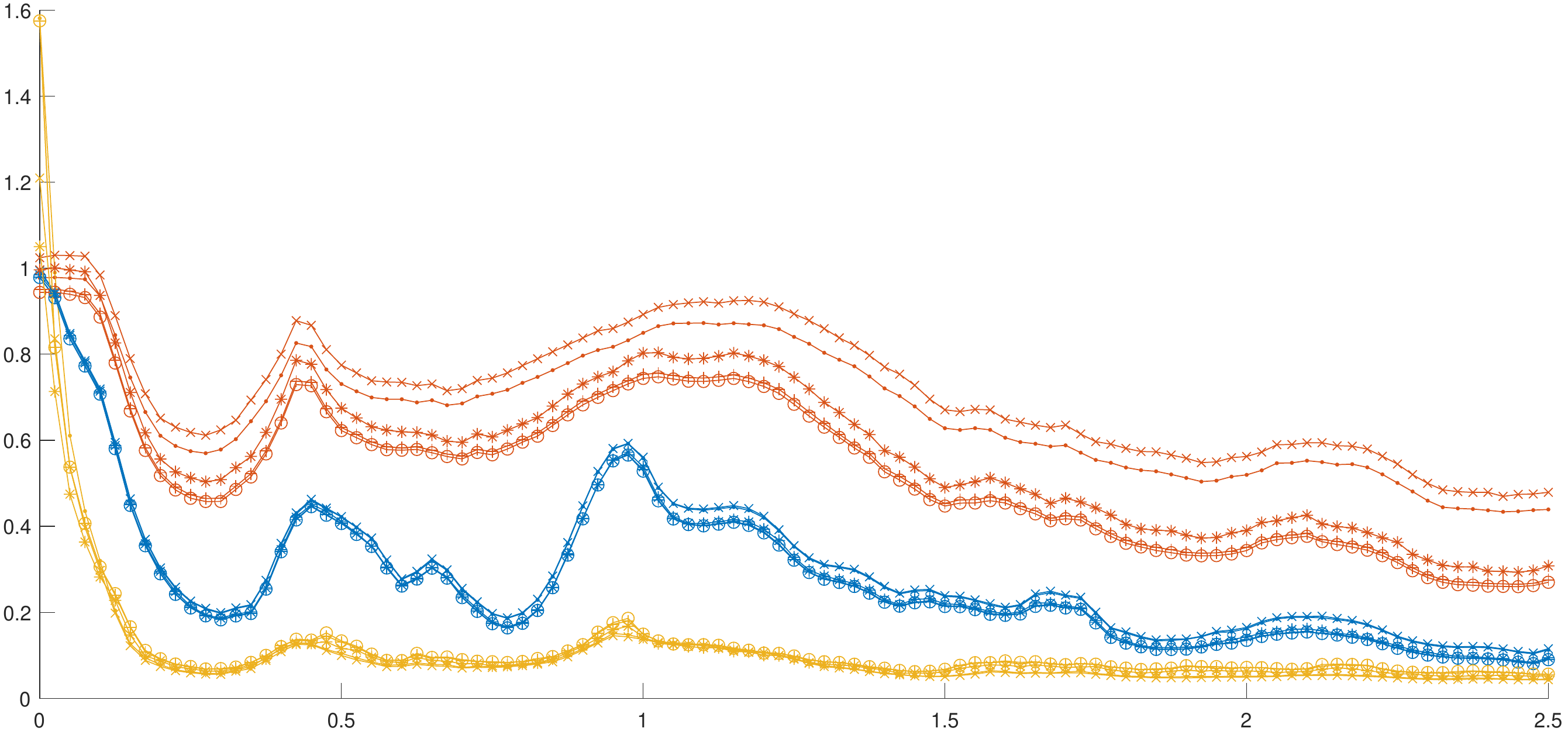}\\
\caption{Different measures of mixing applied to $\bm{v}_{\text{inv}}$, restricted to $A_{3}$, for the double gyre mixer with~$\alpha=0.5$, $\beta=0.5$, and different choices of $\epsilon$.  Red: sample variance; blue: relative mix-norm; yellow: mean length scale. Dots: 8192 boxes on $A_{3}$ and 100 test particles per box; crosses: 8192 boxes and 16 test particles; circles: 2048 boxes and 400 test particles; plus signs: 2048 boxes and 100 test particles; asterisks: 2048 boxes and 16 test particles.}\label{fig:discretize}
\end{center}
\end{figure}

We validated the robustness of the results by studying the influence of the discretization underlying the numerical approximation of the transfer operator. Recall that the transition probabilities (i.e. entries of the transition matrix) are calculated per box. Since particles are assumed to be evenly spread inside a box, diffusion increases with box diameter. We tested different resolutions of the box discretization (Fig.~\ref{fig:discretize}). The graphs of the sample variance, relative mix-norm, and mean length scale remain similar. With increasing box number, meaning a finer discretization, the sample variance moves closer to one since there is less diffusion. The mix-norm and mean length scale appear to be more robust. With increasing number of test particles the sample variance decreases, whereas the mix-norm again remains similar. All in all, the relative mix-norm appears to be the most robust measure with respect to inaccuracies of the numerical approximation of the transfer operator.

\end{document}